\documentclass[preprint,pre,showpacs,preprintnumbers,amsmath,amssymb]{revtex4}

\usepackage{graphicx}
\usepackage{dcolumn}
\usepackage{bm}
\usepackage{braket}
\usepackage{graphicx}
\usepackage{epstopdf}
\usepackage{mwe}    
\usepackage{subfig}
\usepackage[T1]{fontenc}
\usepackage{caption} 
\usepackage{amssymb}
\usepackage{amsmath}
\usepackage{booktabs}
\usepackage{tabularx}
\usepackage{multirow}

\newcommand{\tensr}[1]{\bm{\mathsf{#1}}} 

\usepackage{xcolor}

\begin{document}

\preprint{PREPRINT}

\title{Symmetrized Operator Split Schemes for Force and Source Modeling in Cascaded Lattice Boltzmann Methods for Flow and Scalar Transport}

\author{Farzaneh  Hajabdollahi}
\email{farzaneh.hajabdollahi-ouderji@ucdenver.edu}
\affiliation{Department of Mechanical Engineering, University of Colorado Denver, 1200 Larimer street, Denver, CO  80124, U.S.A.\\}

\author{Kannan N. Premnath}
\email{kannan.premnath@ucdenver.edu}
\affiliation{Department of Mechanical Engineering, University of Colorado Denver, 1200 Larimer street, Denver, CO  80124, U.S.A.\\}


\date{\today}

\begin{abstract}
Operator split forcing schemes exploiting a symmetrization principle, i.e. Strang splitting, for cascaded lattice Boltzmann (LB) methods in two- and three-dimensions for fluid flows with impressed local forces are presented. Analogous scheme for the passive scalar transport represented by a convection-diffusion equation with a source term in a novel cascaded LB formulation is  also derived. They are based on symmetric applications of the split solutions of the changes on the scalar field/fluid momentum due to the sources/forces over half time steps before and after the collision step. The latter step is effectively represented in terms of the post-collision change of moments at zeroth and first orders, respectively, to represent the effect of the sources on the scalar transport and forces on the fluid flow. Such symmetrized operator split cascaded LB schemes are consistent with the second-order Strang splitting and naturally avoid any discrete effects due to forces/sources by appropriately projecting their effects for higher order moments. All the force/source implementation steps are performed only in the moment space and they do not require formulations as extra terms and their additional transformations to the velocity space. These result in particularly simpler and efficient schemes to incorporate forces/sources in the cascaded LB methods unlike those considered previously. Numerical study for various benchmark problems in 2D and 3D for fluid flow problems with body forces and scalar transport with sources demonstrate the validity and accuracy, as well as the second-order convergence rate of the symmetrized operator split forcing/source schemes for the cascaded LB methods.
\end{abstract}

\maketitle
\section{\label{app:Intro}Introduction}

The lattice Boltzmann (LB) method is now a well established alternative numerical technique to computational fluid dynamics (CFD) problems. It derives its basis from kinetic formulations involving the streaming of particle populations along their characteristic directions comprising the lattice, and collisions at lattice nodes represented as a relaxation process, as well as a procedure to represent the effect of impressed forces. The emergent fluid flow behavior is the averaged effect of such stream, collide and forcing steps and thus the LB method may be classified as a mesoscopic approach. Some important advantages of the LB method include its natural framework to incorporate kinetic models for complex flows, ease of implementation of boundary conditions and intrinsic adaptability to parallel computing due to its localized computational steps. As a result, the LB scheme has been successfully applied to a broad range of complex fluid mechanics problems, including multiphase and multicomponent flows, turbulence, thermal convective flows, amongst various other problems~(\cite{Chen1998},~\cite{Succi2001},~\cite{Aidun2010},~\cite{Guo2013}). More recent efforts have focused on further improving the accuracy, stability and efficiency of the LB method to further expand its scope for applications.\par
The collision step, which represents various physics associated with the fluid motion including the momentum diffusion as a relaxation process, plays a main role in the numerical stability of the method. Among the earliest collision models is the single relaxation  time (SRT) model~\cite{Qian1992}, which, while being popular due to its simplicity, is susceptible to numerical instability at relatively high Reynolds numbers. A significant improvement is achieved by the multiple relaxation time model (MRT)~\cite{dHumieres2002} in which different raw moments relax at different rates. More recently, further enhancement in stability was made possible by the introduction of a cascaded LB method, which is a multi-parametric scheme that is based on considering relaxation in terms of central moments, which are formulated by shifting the particle velocity by the local fluid velocity~\cite{Geier2006}. The significant advantages of such more advanced collision models were numerically demonstrated more recently ~\cite{Geier2015}. A strategy to accelerate the convergence of the cascaded LB method has also been devised and studied~\cite{Hajabdollahi2017}, which has been further extended with improved Galilean invariance properties~\cite{hajabdollahi2018galilean}.

Another aspect of the LB schemes, which is particularly important in applications, is the implementation strategy to represent the various impressed body forces, which can either arise within the fluids or imposed externally. Some examples include the local surface tension and phase segregation forces in multiphase fluid systems, Lorentz forces in magnetohydrodynamics, gravity and Coriolis forces. In general, such body forces can be spatially varying and/or time dependent. Due to the kinetic nature of the LB method, special considerations are necessary and various forcing schemes have been introduced over the years~(\cite{He1998b},~\cite{He1999},~\cite{Luo2000},~\cite{Guo2002dis},~\cite{Ladd2001},~\cite{Kupershtokh2004}). In particular, the investigation by~\cite{Guo2002dis} highlighted the discrete effects arising in prior LB forcing schemes via the second order moments in the momentum flux tensor, and provided a consistent source term that avoids such spurious effects when used with the SRT collision model. This was further generalized to the MRT model by including source terms in the moment space in both two-dimensions (2D) and three-dimensions (3D)~(\cite{Mccracken2005},~\cite{Premnath2007},~\cite{Guo2013}).

In the case of the cascaded LB method, the first consistent forcing scheme based on the central moments was presented by~\cite{Premnath2009for}. By taking the source term proposed by~\cite{He1998a} as the starting point, they devised a forcing formulation without discrete effects, which was also shown to be a further generalization of that presented by~\cite{Guo2002dis} to the cascaded LB scheme under appropriate limits. Later,~\cite{Brown2014} constructed another type of forcing scheme for the cascaded LB method based on the exact difference method~\cite{Kupershtokh2004}. More recently,~\cite{Derosis2017a},~\cite{Derosis2017b} and~\cite{fei2017consistent} presented other variants of forcing schemes for LB methods based on central moments. While all these forcing schemes differ from one another due to the variations in the kinetic models for the source term, a common element among them is the presence of extra source terms or changes to the equilibria, which are usually taken together with the collision relaxation terms as part of the collision step. This generally involves computing source moments at different orders and transforming them back to the velocity space, which entails additional computational effort.\par
Based on the consideration that the LB schemes are generally fluid flow, i.e. Navier-Stokes~(NS), solvers, and by avoiding the kinetic aspects for the implementation of the impressed forces, simpler and more efficient strategies can be constructed. The numerical framework for this is the operator splitting approach widely used to efficiently  solve ordinary and partial differential equations arising in various applications including CFD~(\cite{Leveque2002},~\cite{Macnamara2017}). The basis idea is to split the problem into a set of simpler subproblems and then devise a strategy that alternates between solving such simpler problems in certain sequence, which then approximate the solution to the full problem to a certain order of accuracy. Such operator splitting techniques are sometimes also referred to as the fractional step or time-splitting methods. Of particular importance is the Strang spliting~\cite{Strang1968}, which achieves second-order accuracy by a symmetrized application of the solution method for one (or more) of the subproblems. The structure of the higher order splitting errors can be analyzed via the Taylor-Lie series~\cite{Leveque2002} or using the Baker-Compbell-Hausdorff formulas~\cite{Sanz1997}. From such a perspective, Dellar~\cite{Dellar2013} presented a derivation of the lattice Boltzmann method based on Strang splitting with second order accuracy and interpreted both unsplit and time-split forcing schemes based on this approach. In particular, a uncoupled spin-step to implement body force in a SRT LB model introduced earlier by Salmon~\cite{Salmon1999} was shown to be consistent with the Strang splitting. Furthermore, it was also extended to the MRT-LB models~(\cite{Dellar2013},~\cite{Contrino2014}).

In the present investigation, our goal is to construct efficient body force implementation schemes based on the symmetrized operator (Strang) splitting for the cascaded LB methods. The lattice symmetry and the use of central moments naturally impose Galilean invariance for the chosen set of independent moments basis. The symmetric application of the separate body force steps in two half time steps in the cascaded LB formulation provides a second order accuracy. Unlike the unsplit forcing schemes presented earlier for the cascaded LB method~\cite{Premnath2009for}, our approach does not require either the computation of various source moments at different orders or an extra transformation step to convert them back to velocity space. In essence, the operator-split forcing scheme involves one half application of the force before collision and the other half force step after collision. The latter step will be seen to lead to unique expressions for the post-collision change of first order moments in the cascaded collision operator. The precise structure of these expressions will be shown to depend on choice of the first order moment basis vectors associated with the type of lattice considered. In fact, we will present operator split forcing scheme for the cascaded LB method both in 2D and 3D for the computation of the fluid motion. In addition, in order to demonstrate the generality of our approach, we will extend it to represent the convective-diffusion equation (CDE) with a source term, such as those arising in the convective thermal flows with internal heat generation. In this regard, a novel cascaded LB formulation for the solution of the CDE with source term using the Strang splitting will be constructed. Finally, we will present a numerical validation study of the symmetrized operator split forcing/source schemes for the cascaded LB method for fluid flow (i.e., the NS equations) and  passive scalar transport (i.e., the CDE) and in different dimensions.\par
This paper is organized as follows. In the next section (Sec.~2), we briefly review the various operator splitting approaches including the Strang splitting. Section 3 presents the general ideas behind the symmetrized operator splitting based forcing implementation in the LB method. Section 4 discusses the derivation and the algorithmic procedure of the symmetrized operator split forcing scheme for the 2D cascaded LB method for representing fluid flow subjected to local impressed forces. A corresponding 3D formulation is outlined in the Appendix A. Section 5 presents a symmetrized operator split approach source incorporation scheme for a 2D cascaded LB scheme for representing the convection-diffusion based transport of a passive scalar field with local sources. Numerical validation results of various symmetrized operator split forcing/source scheme are presented in Sec.~6. Finally, Sec.~7 summarizes our approach and presents the main conclusion arising from this work.

\section{\label{app:sec2}Operator Splitting Methods}
We will now briefly review the various typical operator splitting methods, including the Strang splitting which will then be exploited to construct efficient second order accurate forcing schemes in the cascaded LB method. For the purpose of illustration, we will consider the numerical solutions of the following evolution problem:
\begin{equation}
\frac{ {d\bm y}}{dt}=\tensr P {\bm y} +\tensr Q {\bm y} ,\quad  {\bm y}(t)={\bm y_0}\quad \text{on} \quad \left[t,t+\Delta t\right],
\label{eq:1}
\end{equation}
where, for ease of presentation, $\tensr P$ and $\tensr Q$ are considered as linear operators. Nonlinear operators can be dealt with using Lie operator formalism~\cite{Sanz1997}. Here, $\Delta t$ is the time step. For reference, the unsplit solution ${\bm y}^{\scriptscriptstyle U}$ of the
full problem can be represented as
 \begin{equation}
{\bm y}^{\scriptscriptstyle U}=e^{\Delta t (\tensr P+\tensr Q)}\bm{y_0}.
\label{eq:2}
\end{equation}
Now, a first order splitting scheme, which is sometimes known as the Lie-Trotter (LT) splitting or as the Godunov splitting scheme in the CFD literature, can be represented by means of the following steps, which compute solution to each subproblem involving $\tensr P$ and $\tensr Q$ separately:
\begin{subequations}
\begin{eqnarray}
 &&\textbf{Step}\, \tensr P:\quad \text{Solve}\quad \dfrac{d\bm {y}^*}{dt'}=\tensr P\bm {y}^* ,\quad  \bm {y}^*(t'=t)=\bm{y_0}\quad \text{on} \quad \left[t,t+\Delta t\right],\\&&
\label{eq:3a}
\textbf{Step}\, \tensr Q:\quad \text{Solve}\quad \dfrac{d\bm {y}^{**}}{dt'}=\tensr Q\bm {y}^{**} ,\quad  \bm {y}^{**}(t'=t)=\bm{y^*}(t+\Delta t)\quad \text{on} \, \left[t,t+\Delta t\right],\label{eq:3b}\\&&
\textbf{Solution}:\quad\bm{y}^{{\scriptscriptstyle LT}}(t+\Delta t)= \bm {y}^{**}(t+\Delta t).\label{eq:3c}
\end{eqnarray}
\end{subequations}
This solution of the Lie-Trotter splitting or the $\tensr P$-$\tensr Q$  splitting scheme may be more compactly represented  by means of the exponential operators as
\begin{equation}
\quad\bm{y}^{{\scriptscriptstyle LT}}(t+\Delta t)= e^{\Delta t \,\tensr Q }e^{\Delta t\, \tensr P }\,\bm{y_0}.
\label{eq:4}
\end{equation}
The local error ($E_l$) incurred over a small time step $\Delta t$ due to splitting when compared to the unsplit solution (Eq.~(\ref{eq:2})) can be estimated by means of a Lie-Taylor series (factored product expansions) as~\cite{Leveque2002}
\begin{equation}
E_{l,\scriptscriptstyle LT}=\bm{y}^{{\scriptscriptstyle LT}}-\bm y^{\scriptscriptstyle U}=\frac{1}{2}\left[\tensr P,\tensr Q\right]\bm{y_0}\Delta t^2+O(\Delta t^3),
\label{eq:5}
\end{equation}
where the symbol $\left[\mathord{\cdot} , \mathord{\cdot} \right]$ represents the commutator, i.e., $\left[\tensr{X} , \tensr{Y}\right]=\tensr{X}\tensr{Y}-\tensr{Y}\tensr{X}$ for any two operators $\tensr{X}$ and $\tensr{Y}$. Then, the global error ($E_g$) over a time duration $T$ or $T/\Delta t$ number of steps is $E_{g,\scriptscriptstyle LT}=(T/\Delta t)\cdot E_{l,\scriptscriptstyle LT}\sim O(\Delta t)$, which means that the Lie-Trotter scheme is first order accurate. This means that even if a higher order method is used to solve each subproblem ($\textbf{Step}\, \tensr P$ and $\textbf{Step}\, \tensr Q$), the above splitting scheme is still overall first order accurate due to the decomposition error arising from the non-commuting operators, which is often the case in practice.\par
One possibility to improve the order of  accuracy is to symmetrize the computation via taking the average of the two sequences of calculations, i.e. $\textbf{Step}\, \tensr P$ - $\textbf{Step}\, \tensr Q$ and $\textbf{Step}\, \tensr Q$ - $\textbf{Step}\, \tensr P$ results. Such an averaged scheme may be represented as~\cite{strang1963accurate}
  \begin{equation}
\quad\bm{y}^{{\scriptscriptstyle A}}= \frac{1}{2}(e^{\Delta t \,\tensr P }e^{\Delta t\, \tensr Q }+e^{\Delta t \,\tensr Q }e^{\Delta t\, \tensr P })\bm{y_0}.
\label{eq:6}
\end{equation}
This approach introduces a local error relative to the unsplit solution  (Eq.~(\ref{eq:2})), which can be written as~\cite{hundsdorfer2013numerical}
\begin{equation*}
E_{l,\scriptscriptstyle A}=\bm{y}^{{\scriptscriptstyle A}}-\bm y^{\scriptscriptstyle U}=\tensr R'\Delta t^3+O(\Delta t^4),
\end{equation*}
where
\begin{equation*}
\tensr R'=-\frac{1}{12}(\left[\tensr P,\left[\tensr P,\tensr Q\right]\right]+\left[\tensr Q,\left[\tensr Q,\tensr P\right]\right])\bm{y_0}.
\end{equation*}
Hence, the global error becomes $E_{g,\scriptscriptstyle A}=(T/\Delta t)\cdot E_{l,\scriptscriptstyle A}\sim O(\Delta t^2)$. While this is theoretically interesting to gain an order of accuracy, it is computationally expensive as, for each time step, double the effort is required when compared to the previous scheme ($\tensr P-\tensr Q$ splitting).\par
A more efficient strategy to achieve a global second order  accuracy is to devise the Strang (S) splitting ~\cite{Strang1968}. In this scheme, one of the operators (say $\tensr P$) is applied twice for a time step of length $\Delta t/2$, before and after the solution of the other subproblem (say, involving $\textbf{Step}\, \tensr Q$), which is solved for full step length of $\Delta t$. This may be represented as
\begin{subequations}
\begin{eqnarray}
 &&\textbf{Step}\, \tensr P^{1/2}:\quad \text{Solve}\quad \dfrac{d\bm {y}^*}{dt'}=\tensr P\bm {y}^* ,\quad  \bm {y}^*(t'=t)=\bm{y_0}\quad \text{on} \quad \left[t,t+\Delta t/2\right],\\
\label{eq:7a}
&&\textbf{Step}\, \tensr Q:\quad \text{Solve}\quad \dfrac{d\bm {y}^{**}}{dt'}=\tensr Q\bm {y}^{**} ,\quad  \bm {y}^{**}(t'=t)=\bm{y^*}(t+\Delta/2)\quad \text{on} \, \left[t,t+\Delta t\right],\\
\label{eq:7b}
&&\textbf{Step}\, \tensr P^{1/2}:\quad \text{Solve}\quad \dfrac{d\bm {y}^{***}}{dt'}=\tensr P\bm {y}^{***} ,\quad  \bm {y}^{***}(t'=t)=\bm {y}^{**}(t+\Delta t)\quad \text{on} \quad \left[t,t+\Delta t/2\right],\\
\label{eq:7c}
 && \textbf{Solution}:\quad\bm{y}^{{\scriptscriptstyle S}}(t+\Delta t)= \bm {y}^{***}(t+\Delta t/2).\qquad\qquad
\label{eq:7d}
\end{eqnarray}
\end{subequations}
This symmetric application of the operators in the $\tensr P^{1/2}-\tensr Q-\tensr P^{1/2}$ scheme achieves second order accuracy, which may be deduced by first noting that the Strang splitting solution may be more compactly written in the exponential form as
\begin{equation}
\quad\bm{y}^{{\scriptscriptstyle S}}(t+\Delta t)= e^{\Delta t/2 \,\tensr P }\,e^{\Delta t\, \tensr Q }\,e^{\Delta t/2 \,\tensr P }\bm{y_0}.
\label{eq:8}
\end{equation}
Its local error when compared to the unsplit solution  (Eq.~(\ref{eq:2})) then follows via a Lie-Taylor series as~\cite{Macnamara2017}
\begin{equation}
E_{l,\scriptscriptstyle S}=\bm{y}^{{\scriptscriptstyle S}}-\bm y^{\scriptscriptstyle U}=\tensr R\Delta t^3+O(\Delta t^4),
\label{eq:9}
\end{equation}
where
\begin{equation}
\tensr R=\frac{1}{24}(\left[\left[\tensr P,\tensr Q\right],\tensr P\right]+2\left[\left[\tensr P,\tensr Q\right],\tensr Q\right])\bm{y_0}.
\label{eq:10}
\end{equation}
Then, the global error ($E_g$) over a time period $T$ follows as $E_{g,\scriptscriptstyle S}=(T/\Delta t)\cdot E_{l,\scriptscriptstyle S}\sim O(\Delta t^2)$ and hence this scheme is second order accurate. An equally valid possibility to achieve a similar second order accuracy is to consider the $\tensr Q^{1/2}-\tensr P-\tensr Q^{1/2}$ splitting, which is useful when  $\textbf{Step}\, \tensr P$ is more expensive to compute than $\textbf{Step}\, \tensr Q$. It may be noted that a similar scheme was independently devised by~\cite{Marchuk1968}, who further analyzed and elaborated on its variants (see also~\cite{marchuk1975methods}), and hence it is sometimes referred to as the Strang-Marchuk splitting scheme.

\section{\label{app:Sec3}Strang Splitting of Lattice Boltzmann Method Including Body Forces}

Lattice Boltzmann (LB) schemes are generally constructed to represent the evolution of the dynamics of the fluid motion represented by
\begin{subequations}
\begin{eqnarray}
&\partial_t \rho +\bm \nabla \cdot (\rho \bm u)=0,\label{eq:11a}\\
&\partial_t(\rho \bm u)+\bm \nabla \cdot(\rho \bm u \bm u)=-\bm \nabla P+\bm \nabla \cdot \tensr \Pi_{\scriptscriptstyle V}+\bm F,\label{eq:11b}
\end{eqnarray}
\end{subequations}
where $\rho$ and $\bm u$ are the fluid density and velocity, respectively, $P$ is the pressure and $\tensr \Pi_{\scriptscriptstyle V}$ is the viscous stress tensor. Here, $\bm F$ represents the effect of the local impressed body forces, which can vary spatially and may be time dependent, i.e. for e.g. in 2D, $\bm F=(F_x,F_y)$ where $F_x=F_x(\bm x,t)$ and $F_y=F_y(\bm x,t)$ . An efficient approach to solve the above fluid flow equation in the LB framework is to solve the Eqs.~(\ref{eq:11a}) and~(\ref{eq:11b}), but without the body force $\bm F$ using the usual stream and collide procedure (subproblem A) and then separately solve $\partial_t(\rho \bm u)=\bm F$ as a forcing step (subproblem B) and subsequently combined appropriately in a certain sequence to yield a second order accurate scheme. This can be achieved via symmetrization of the operator splitting of the one of the subproblems over two half time steps. Dellar ~\cite{Dellar2013} performed a derivation and analysis of the LB method via Strang splitting, which will be used as formal starting point to construct efficient operator split forcing schemes for the cascaded LB method in the subsequent sections.\par
In the following, $\tensr S$, $\tensr C$ and $\tensr F$ are used to denote the operators used to perform the streaming step, collision step and the forcing step, respectively. For a lattice containing $\alpha=0,1,2,\dots b$ directions, the collision and streaming steps can be represented as
\begin{subequations}
\begin{eqnarray}
 &&\textbf{Step}\, \tensr C:\quad \mathbf{f}(\bm x,t+\Delta t)=\tensr C \mathbf f(\bm x, t)=\mathbf f(\bm x, t)+\tensr K \cdot \mathbf {\widehat{g}},
\label{eq:12a}\\
&&\textbf{Step}\, \tensr S:\quad {f_{\alpha}}(\bm x,t+\Delta t)=\tensr S  {f_{\alpha}}(\bm x, t)\equiv {f_{\alpha}}(\bm x-\mathbf e_{\alpha}\Delta t,t).
\label{eq:12b}
\end{eqnarray}
\end{subequations}
Here, $\mathbf f=(f_0,f_1,f_2\dots f_b)^\dagger$ is a vector of size ($b+1$) representing the distribution functions, where $\dag$ is the transpose operator, $\mathbf {\widehat g}=(\widehat g_0,\widehat g_1,\widehat g_2\dots \widehat g_b)^\dag$  is the vector representing the change of different moments under collision, and $\tensr K$ is the transformation matrix of the cascaded LB method that maps changes in moments back to changes in the distribution functions, which are specified later.

It may be noted that $\tensr C $ and $\tensr S$ operators represent the split solution operators of the discrete analog of $\partial_t f_{\alpha}=\Omega_{\alpha}$ and  $\partial_t f_{\alpha}+\mathbf {e_{\alpha}}\cdot{\bf \nabla} { f_{\alpha}}=0$, respectively, of the discrete velocity Boltzmann equation $\partial_t  f_{\alpha}+\mathbf e_{\alpha} \cdot{\bf \nabla} {f_{\alpha}}=\Omega_{\alpha}$, whose emergent behavior represents the NS equations given in Eq.~(\ref{eq:11a}) and Eq.~(\ref{eq:11b}), but without $\bm F$. Then, the forcing step separately solves the following:
\begin{equation}
\textbf{Step}\,  \tensr F: \frac{\partial}{\partial t}(\rho \bm u)=\bm {F}.
\label{eq:13}
\end{equation}
One possibility to combine the above split steps to effectively achieve second order accuracy is to perform a symmetric application of the forcing steps over two half time steps, before and after the collision step, which is akin to the spin steps for the force presented by Salmon~(\cite{Salmon1999}):

\begin{eqnarray}
 {f_{\alpha}}(\bm x,t+\Delta t)=\tensr S\, \tensr F^{1/2}\,\tensr C\, \tensr F^{1/2}{f_{\alpha}}(\bm x,t),
\label{eq:14}
\end{eqnarray}
where $\tensr F^{1/2}$ represents performing the solution of Eq.~(\ref{eq:13}) over time step of length $\Delta t/2$. Ref.~\cite{Dellar2013} showed that this achieves second order accuracy similar to the Strang splitting extended to three operators: ${{f^{'}}_{\alpha}}(\bm x,t+\Delta t)=\tensr C^{1/2}\tensr F^{1/2}\,\tensr S\, \tensr F^{1/2} \tensr C^{1/2}{{f^{'}}_{\alpha}}(\bm x,t)$, where the two are related by $ f^{'}_{\alpha}=\tensr C^{1/2} \tensr F^{1/2} f_{\alpha}$ . Since the momentum is conserved during collisions, a second order scheme with Eq.~(\ref{eq:14}) can be obtained by $\rho \bm u=\sum_{\alpha} {f^{'}}_{\alpha} \mathbf e_{\alpha}=\tensr F^{1/2}(\sum_{\alpha} f_{\alpha} \mathbf e_{\alpha})$. We will adopt the above strategy in our derivation of the symmetrized operator split forcing scheme for the cascaded LB method in the subsequent sections. Similar approach was recently adopted for the MRT LB models~(e.g.,~\cite{Contrino2014}). In addition, Schiller~\cite{schiller2014unified} proposed a variant of the Strang splitting of forcing steps around streaming and collisions, where the half collision step is valid for the regime involving the relaxation time being much greater than the time step. Also, Dellar~\cite{Dellar2014} showed that the Crank-Nicolson solution of the moment equations for combined collisions and time-independent forcing obtained by Strang splitting is equivalent to Kupershtokh's exact difference method~\cite{Kupershtokh2004}.

\section{\label{app:d3q27matrix}Body Force Scheme for 2D Cascaded LB Method for Fluid Flow via Strang Splitting}

We will consider a 2D cascaded LB formulation for a two-dimensional, nine velocity (D2Q9) lattice. The components of the particle velocities are then represented by the following vectors using the standard Dirac's bra-ket notation:
\begin{subequations}
\begin{eqnarray}
&\ket{e_{x}} =\left(     0,     1,    0,     -1,     0,  1, -1, -1,  1  \right)^\dag,\label{eq:15a}\\
&\ket{e_{y}} =\left(     0,     0,     1,     0,    -1,  1,  1, -1, -1
\right)^\dag.\label{eq:15b}
\end{eqnarray}
\end{subequations}
Their components for any particle velocity direction $\alpha$ (where $\alpha=0,\ldots, 8$) are referred to as $e_{\alpha x}$ and $e_{\alpha y}$, respectively. Furthermore, we need the following 9-dimensional vector:
\begin{eqnarray}
&\ket{1} =\left(     1,     1,    1,     1,     1,  1,  1,  1,  1  \right)^\dag.\label{eq:16}
\end{eqnarray}
The zeroth moment is the Euclidean inner product of this vector with the distribution function.
We then consider the following specific set of orthogonal basis vectors used in the collision term of the cascaded LB method (e.g.,~\cite{Premnath2009for}):
\begin{eqnarray}
\ket{K_0}=\ket{1}, \quad \ket{K_1}=\ket{e_{x}}, \quad \ket{K_2}=\ket{e_{y}}, \quad \ket{K_3}=3\ket{e_{x}^2+e_{y}^2}-4\ket{1},\nonumber \\
\ket{K_4}=\ket{e_{x}^2-e_{y}^2}, \quad
\ket{K_5}=\ket{e_{x}e_{y}}, \quad
\ket{K_6}=-3\ket{e_{x}^2e_{y}}+2\ket{e_{y}},\nonumber \\
\ket{K_7}=-3\ket{e_{x}e_{y}^2}+2\ket{e_{x}}, \quad
\ket{K_8}=9\ket{e_{x}^2e_{y}^2}-6\ket{e_{x}^2+e_{y}^2}+4\ket{1}.\label{eq:17}
\end{eqnarray}
In the above, symbol such as $\ket{e_x^2e_y}=\ket{e_xe_xe_y}$ represents a vector resulting from the elementwise vector multiplication (Hadamard product) of the sequence of vectors $\ket{e_x}$, $\ket{e_x}$ and $\ket{e_y}$. By combining the above $9$ vectors, we then obtain the following orthogonal matrix
\begin{equation}
\tensr{K}=\left[\ket{K_0},\ket{K_1},\ket{K_2},\ket{K_3},\ket{K_4},\ket{K_5},\ket{K_6},\ket{K_7},\ket{K_8}\right].
\label{eq:18}
\end{equation}
Here, $\tensr{K}$ maps changes of moments under collisions back to changes in the distribution functions.
In order to determine the structure of the cascaded collision operator, we first define the following set of central moments of the distribution functions and its equilibria of order ($m+n$), respectively, as
\begin{eqnarray}
\left( {\begin{array}{*{20}{l}}
{{{\hat \kappa }_{{x^m}{y^n}}}}\\
{\hat {\kappa} _{{x^m}{y^n}}^{eq}}
\end{array}} \right) = \sum\limits_\alpha  {\left( {\begin{array}{*{20}{l}}
{{f_\alpha }}\\
{f_\alpha ^{eq}}
\end{array}} \right)} {{(e_{\alpha x}-u_x)}^m}{{(e_{\alpha y}-u_y)}^n}.
\label{eq:19}
\end{eqnarray}
By equating the discrete central moments of the equilibrium distribution function with the corresponding continuous central moments based on the local Maxwellian~(\cite{Geier2006},~\cite{Asinari2008}), we get
\begin{eqnarray}
&\widehat{\kappa}^{eq}_{0}=\rho,\,
\widehat{\kappa}^{eq}_{x}=0,\,
\widehat{\kappa}^{eq}_{y}=0,\,
\widehat{\kappa}^{eq}_{xx}=c_s^2\rho,\,
\widehat{\kappa}^{eq}_{yy}=c_s^2\rho,\nonumber \\
&\widehat{\kappa}^{eq}_{xy}=0,\,
\widehat{\kappa}^{eq}_{xxy}=0,\,
\widehat{\kappa}^{eq}_{xyy}=0,\,
\widehat{\kappa}^{eq}_{xxyy}=c_s^4\rho.
\label{eq:20}
\end{eqnarray}
where $c^2_s=1/3$ with $c_s$ being the sound speed. This is set by applying the usual lattice units, i.e. $\Delta x = \Delta t = 1$ or the particle speed $c=\Delta x/\Delta t=1$, and because $c_s^2=c^2/3$ for the athermal LB scheme used in this work (see e.g.~\cite{Kruger2016}). On the other hand, the actual computations in the cascaded formulations are carried out in terms of raw moments, which are defined as (designated here with the ($'$) symbol)
\begin{eqnarray}
\left( {\begin{array}{*{20}{l}}
{{{\hat \kappa }_{{x^m}{y^n}}}}^{'}\\
{\hat {\kappa} _{{x^m}{y^n}}^{eq'}}
\end{array}} \right) = \sum\limits_\alpha  {\left( {\begin{array}{*{20}{l}}
{{f_\alpha }}\\
{f_\alpha ^{eq}}
\end{array}} \right)} {e_{\alpha x}^m}{e_{\alpha y}^n}.
\label{eq:21}
\end{eqnarray}
The collide and stream steps ($\tensr C$ and $\tensr S$) of the 2D cascaded LB method can then be, respectively, written as~\cite{Geier2006}
\begin{subequations}
\begin{eqnarray}
 &\textbf{Step}\, \tensr C:\quad f^{p}_{\alpha}=f_{\alpha}+(\tensr K\cdot \widehat {\mathbf g})_{\alpha}
\label{eq:22a}\\
&\textbf{Step}\, \tensr S:\quad f_{\alpha}(\bm x, t)=f^{p}_{\alpha}(\bm x-\bm e_{\alpha}\Delta t,t),
\label{eq:22b}
\end{eqnarray}
\end{subequations}
where $f^{p}_{\alpha}$ represents the post-collision distribution function and $\mathbf {\widehat g}=(\widehat g_0,\widehat g_1,\widehat g_2\dots \widehat g_8)^\dag$ is the change of different moments under collisions, which is determined based on the relaxation of various central moments to their corresponding equilibria in a cascaded fashion~\cite{Geier2006}. Since the mass and momentum are collision invariants, $\widehat g_0=\widehat g_1=\widehat g_2=0$. As a result, the cascaded structure starts from the non-conserved second order moments, and the corresponding components of the change of different moments under collisions are given by
\begin{align}
\widehat{g}_3&=\frac{\omega_3}{12}\left\{ \frac{2}{3}\rho+{\rho(u_x^2+u_y^2)}
-(\widehat{{\kappa}}_{xx}^{'}+\widehat{{\kappa}}_{yy}^{'})
\right\}, \nonumber \\
\widehat{g}_4&=\frac{\omega_4}{4}\left\{{\rho(u_x^2-u_y^2)}
-(\widehat{{\kappa}}_{xx}^{'}-\widehat{{\kappa}}_{yy}^{'})
\right\}, \nonumber \\
\widehat{g}_5&=\frac{\omega_5}{4}\left\{{\rho u_x u_y}
-\widehat{{\kappa}}_{xy}^{'}
\right\},\nonumber \\
\widehat{g}_6&=\frac{\omega_6}{4}\left\{2\rho u_x^2 u_y+\widehat{{\kappa}}_{xxy}^{'}
              -2u_x\widehat{{\kappa}}_{xy}^{'}-u_y\widehat{{\kappa}}_{xx}^{'}
              \right\}-\frac{1}{2}u_y(3\widehat{g}_3+\widehat{g}_4)-2u_x\widehat{g}_5,\nonumber \\
\widehat{g}_7&=\frac{\omega_7}{4}\left\{2\rho u_x u_y^2+\widehat{{\kappa}}_{xyy}^{'}
              -2u_y\widehat{{\kappa}}_{xy}^{'}-u_x\widehat{{\kappa}}_{yy}^{'}
              \right\}-\frac{1}{2}u_x(3\widehat{g}_3-\widehat{g}_4)-2u_y\widehat{g}_5,\nonumber \\
\widehat{g}_8&=\frac{\omega_8}{4}\left\{\frac{1}{9}\rho+3\rho u_x^2 u_y^2-\left[\widehat{{\kappa}}_{xxyy}^{'}
                                 -2u_x\widehat{{\kappa}}_{xyy}^{'}-2u_y\widehat{{\kappa}}_{xxy}^{'}
                                 +u_x^2\widehat{{\kappa}}_{yy}^{'}+u_y^2\widehat{{\kappa}}_{xx}^{'}\right.\right.
                                 \nonumber \\
                                 &\left.\left.+4u_xu_y\widehat{{\kappa}}_{xy}^{'}
                                 \right]
                                  \right\}-2\widehat{g}_3-\frac{1}{2}u_y^2(3\widehat{g}_3+\widehat{g}_4)
                                  -\frac{1}{2}u_x^2(3\widehat{g}_3-\widehat{g}_4)\nonumber\\
                                  &-4u_xu_y\widehat{g}_5-2u_y\widehat{g}_6
                                  -2u_x\widehat{g}_7.\label{eq:23}
\end{align}
where $\omega_3, \omega_4,\ldots, \omega_8$ are the relaxation parameters. These relaxation steps lead to the following expressions for the bulk and shear viscosities, respectively, as $\zeta=\frac{1}{3}(\frac{1}{\omega_3}-\frac{1}{2})\Delta t$ and $\nu=\frac{1}{3}(\frac{1}{\omega_j}-\frac{1}{2})\Delta t$ where $j=4,5$, and the pressure field $P$ is obtained via an equation of state as $P=\frac{1}{3}\rho$.\par
After the streaming step, i.e., Eq.~(\ref{eq:22b}), we obtain the output velocity field components (designated with a superscript ${}" o"$) as the first moment of $f_{\alpha}$:
\begin{eqnarray}
 &\rho u^o_x=\sum_{\alpha=0}^8 {f}_{\alpha}  e_{\alpha x},\quad  \rho u^o_y=\sum_{\alpha=0}^8 {f}_{\alpha} e_{\alpha y}.
\label{eq:24}
\end{eqnarray}
We then introduce the effect of the body force $\bm F=(F_x,F_y)$ as a solution of the subproblem in  Eq.~(\ref{eq:13}). This is accomplished by performing two symmetric steps of half time steps of length $\Delta t/2$, one before and the other after the collision step. Both these steps incorporate the effect of forces directly into the moment space. Solving  Eq.~(\ref{eq:13}) for the first part of the symmetric sequence of step yields $\rho u_x-\rho u^o_x =F_x\frac{\Delta t}{2}$ and $\rho u_y-\rho u^o_y =F_y\frac{\Delta t}{2}$. Thus,
\begin{equation}
\text{Pre-collision Forcing Step}\,  \tensr F^{1/2}: u_x=\frac{1}{\rho}\left(\rho u^o_x+\frac{F_x}{2}\Delta t\right),\quad u_y=\frac{1}{\rho}\left(\rho u^o_y+\frac{F_y}{2}\Delta t\right).
\label{eq:25}
\end{equation}
Then, we use this updated velocity field $(u_x,u_y)$ in Eq.~(\ref{eq:23}) to perform the cascaded relaxation collision step to determine the change of different moments under collisions, i.e. $\widehat g_{\beta}$, $\beta=3,4,\dots,8$. As a result of correctly projecting the effect of the forces in the various higher order moments, it naturally eliminates the discrete effects identified earlier~\cite{Guo2002dis} (see the discussion at the end of this section). Then, to implement the other part of the symmetrized force step with half step to solve Eq.~(\ref{eq:13}) post collision, we set $\rho u^p_x-u_x=F_x\frac{\Delta t}{2}$ and $\rho u^p_y-u_y=F_y\frac{\Delta t}{2}$ , where $(u^p_x,u^p_y)$ is the result of the target velocity field due to the forcing step after collision. Thus,
\begin{equation}
\text{Post-collision Forcing Step}\,  \tensr F^{1/2}: \rho u^p_x=\rho u_x+\frac{F_x}{2}\Delta t,\quad \rho u^p_y=\rho u_y+\frac{F_y}{2}\Delta t.
\label{eq:26}
\end{equation}
Note that this can also be rewritten in terms of the output velocity field  $\bm u^o=(u^o_x,u^o_y)$ given in Eq.~(\ref{eq:24}) by using Eq.~(\ref{eq:25}) as
\begin{equation}
 \rho u^p_x=\rho u^o_x+{F_x}\Delta t,\quad \rho u^p_y=\rho u^o_y+{F_y}\Delta t.
\label{eq:27}
\end{equation}
A main issue here is how to effectively design the post-collision distribution function $f^p_{\alpha}$ in the cascaded LB method so that Eq.~(\ref{eq:27}) is precisely satisfied. Now, using $ f^{p}_{\alpha}=f_{\alpha}+(\tensr K\cdot \widehat {\mathbf g})_{\alpha}$ and taking its first moments, we get
\begin{subequations}
\begin{eqnarray}
 &\rho u^p_x=\Sigma_{\alpha}f^p_{\alpha}e_{\alpha x}=\Sigma_{\alpha}f_{\alpha}e_{\alpha x}+\Sigma_{\beta}{\braket{{K_{\beta}}|{e_{x}}}} \widehat g_{\beta},
\label{eq:28a}\\
 &\rho u^p_y=\Sigma_{\alpha}f^p_{\alpha}e_{\alpha y}=\Sigma_{\alpha}f_{\alpha}e_{\alpha y}+\Sigma_{\beta}{\braket{{K_{\beta}}|{e_{y}}}} \widehat g_{\beta}.
\label{eq:28b}
\end{eqnarray}
\end{subequations}
Based on the orthogonal basis vectors $\ket{K_{\beta}}$ given in Eq.~(\ref{eq:17}), it follows that
\begin{equation}
 \Sigma_{\beta}{\braket{{K_{\beta}}|{e_{x}}}}g_{\beta}=6 \widehat {g}_1,\quad \Sigma_{\beta}{\braket{{K_{\beta}}|{e_{y}}}}g_{\beta}=6 \widehat {g}_2.
\label{eq:29}
\end{equation}
Using Eqs.~(\ref{eq:24}) and~(\ref{eq:29}) in Eqs.~(\ref{eq:28a}) and~(\ref{eq:28b}) we, get the desired velocity field as
\begin{equation}
\rho u^p_x=\rho u^o_x+6 \widehat {g}_1, \quad \rho u^p_y=\rho u^o_y+6 \widehat {g}_2.
\label{eq:30}
\end{equation}
Comparing the result of the target velocity field following the second half of the symmetrized forcing steps given in Eq.~(\ref{eq:27}) with the change of moments based expressions in Eq.~(\ref{eq:30}), we obtain
\begin{equation}
\widehat {g}_1=\frac{F_x}{6}\Delta t, \quad \widehat {g}_2=\frac{F_y}{6}\Delta t.
\label{eq:31}
\end{equation}
Equation~(\ref{eq:31}) represents an algorithmic result that effectively implements the effect of the post-collision forcing step over a duration of half time step following collision. This is a consequence of the momentum needing to change by $\bm{F}\Delta t$ over a time step, and the normalization is implied by our choice of basis for the moments.
Then, the above relation (Eq.~(\ref{eq:31})) for the post-collision change of first moments due to the force field ($\widehat{g}_1$ and $\widehat{g}_2$) along with the change of different higher moments under collisions $\widehat{g}_\beta$, where $\beta=3,4,\ldots, 8$, given in Eq.~(\ref{eq:23}) effectively provide the desired post-collision states of the distribution function $f^p_{\alpha}$. Expanding Eq.~(\ref{eq:22a}), we get the expressions for the post-collision distribution functions as
\begin{eqnarray}\nonumber
f^p_0&=&{f}_{0}+\left[\widehat{g}_0-4(\widehat{g}_3-\widehat{g}_8)\right], \nonumber \\
f^p_1&=&{f}_{1}+\left[\widehat{g}_0+\widehat{g}_1-\widehat{g}_3+\widehat{g}_4    +2(\widehat{g}_7-\widehat{g}_8)\right], \nonumber \\
f^p_2&=&{f}_{2}+\left[\widehat{g}_0+\widehat{g}_2-\widehat{g}_3-\widehat{g}_4
+2(\widehat{g}_6-\widehat{g}_8)\right], \nonumber \\
f^p_3&=&{f}_{3}+\left[\widehat{g}_0-\widehat{g}_1-\widehat{g}_3+\widehat{g}_4
-2(\widehat{g}_7+\widehat{g}_8)\right],\nonumber \\
f^p_4&=&{f}_{4}+\left[\widehat{g}_0-\widehat{g}_2-\widehat{g}_3-\widehat{g}_4
-2(\widehat{g}_6+\widehat{g}_8)\right], \nonumber \\
f^p_5&=&{f}_{5}+\left[\widehat{g}_0+\widehat{g}_1+\widehat{g}_2+2\widehat{g}_3
+\widehat{g}_5-\widehat{g}_6-\widehat{g}_7+\widehat{g}_8\right], \nonumber \\
f^p_6&=&{f}_{6}+\left[\widehat{g}_0-\widehat{g}_1+\widehat{g}_2+2\widehat{g}_3
-\widehat{g}_5-\widehat{g}_6+\widehat{g}_7+\widehat{g}_8\right], \nonumber \\
f^p_7&=&{f}_{7}+\left[\widehat{g}_0-\widehat{g}_1-\widehat{g}_2+2\widehat{g}_3
+\widehat{g}_5+\widehat{g}_6+\widehat{g}_7+\widehat{g}_8\right], \nonumber \\
f^p_8&=&{f}_{8}+\left[\widehat{g}_0+\widehat{g}_1-\widehat{g}_2+2\widehat{g}_3
-\widehat{g}_5+\widehat{g}_6-\widehat{g}_7+\widehat{g}_8\right]. \label{eq:32}
\end{eqnarray}

Then, the algorithmic procedure of our symmetrized operator split forcing scheme for the 2D cascaded method can be summarized in terms of the following sequence of steps to evolve for a time duration $[t,t+\Delta t]$:
%
%
%
%
%
\renewcommand{\theenumi}{(\roman{enumi})} 
\begin{enumerate}
\item Obtain the updated the velocity $\bm u=(u_x,u_y)$ based on the pre-collision forcing with half step using Eq.~(\ref{eq:25}).

\item Compute the change of moments under collisions, $\widehat{g}_{\beta}$, $\beta=3,4,\dots,8$ using Eq.~(\ref{eq:23}) based on the updated velocity $(u_x,u_y)$ obtained in Step (i).

\item Perform post-collision forcing with a half step effectively via the calculation of change of first order moments, i.e. $\widehat{g}_1$ and $\widehat{g}_2$ using Eq.~(\ref{eq:31}).

\item Compute the post-collision distribution functions  $f^p_{\alpha}$ , $\alpha=0,1,\dots,8$ using Eq.~(\ref{eq:32}).

\item Perform the streaming step using Eq.~(\ref{eq:22b}) to obtain the updated distribution functions $f_{\alpha}$ , $\alpha=0,1,\dots,8$.

\item Finally, obtain the output velocity field $\bm u^o=(u^o_x,u^o_y)$ via Eq.~(\ref{eq:24}) and the density $\rho$ using $\rho=\sum_{\alpha=0}^8f_{\alpha}$.
\end{enumerate}
Some of the main advantages of this symmetrized operator split forcing scheme of the cascaded LB method are:
%
\renewcommand{\theenumi}{(\alph{enumi})} 
\begin{enumerate}
 \item Using symmetrization principle with half-time step application of the body force before and after collision is consistent with Strang splitting and the scheme is formally second order accurate in time.

\item The approach correctly projects the effects of the body force on the higher order moments via step (ii) above and hence naturally eliminates the discrete effects identified in prior works ~\cite{Guo2002dis} (see below for details).

\item The procedure is simple and efficient by involving the body force implementation directly only in the moment space and does not require additional terms due to forcing in the velocity space, which is usually obtained via cumbersome transformation from the moment space as in prior forcing schemes for the cascaded LB method. This aspect is especially advantageous in 3D. Appendix A outlines the implementation of this approach for a 3D central moment based LB scheme.
\end{enumerate}

We will now present an analysis on how the spurious term $F_iu_j+F_ju_i$ that can appear in the viscous stress is eliminated in our present central moments-based cascaded LB formulation using a split force implementation. This can be achieved by a continuous time equation for the second central moment whose evolution is independent of the body force. As a result, it can introduce a canceling second moment of the body force term at the leading order in the emergent PDE of the second raw moment of the distribution functions recovering correct flow physics. We will start with this latter aspect first and identify this compensating second raw moment of the body force by considering the discrete velocity Boltzmann equation $\partial_t  f_{\alpha}+\mathbf e_{\alpha} \cdot{\bf \nabla} {f_{\alpha}}=\Omega_{\alpha}+S_{\alpha}$, where $\Omega_{\alpha}$ and $S_{\alpha}$ are the collision operator and the source term due to the body force, respectively. Taking its zeroth and first moments lead to
\begin{equation}
\partial_t \rho + \bm{\nabla}\cdot(\rho \bm{u}) =0, \quad \partial_t (\rho\bm{u}) + \bm{\nabla}\cdot \tensr{\Gamma} =\bm{F}, \label{eq:zeroth_first_moment}
\end{equation}
and then taking its second moment, we obtain the following evolution equation
\begin{equation}
\partial_t \tensr{\Gamma} + \bm{\nabla}\cdot \tensr{\Lambda} =-\frac{1}{\tau} \tensr{\Gamma}^{(neq)}+\tensr{\Upsilon}, \label{eq:second_moment}
\end{equation}
where $\tensr{\Gamma}$ and $\tensr{\Lambda}$ are the second and third moments of the distribution functions, i.e., $\sum_\alpha f_\alpha e_{\alpha i} e_{\alpha j}$ and  $\sum_\alpha f_\alpha e_{\alpha i} e_{\alpha j} e_{\alpha k}$, respectively, and $\tensr{\Upsilon}$ is the required canceling second
moment of the body force term, i.e., $\sum_\alpha S_\alpha e_{\alpha i} e_{\alpha j}$, which should arise via a condition on the second central moment given in the following. In Eq.~(\ref{eq:second_moment}), $\tensr{\Gamma}^{(neq)}$ is the non-equilibrium part of the second raw moment and $\tau=1/\omega_j$, where $j=4,5$, is the corresponding relaxation time, which are related to the viscous stress.

In order to determine the evolution equation for hydrodynamics at the leading order, we now apply the Chapman-Enskog (C-E) expansions of the distribution functions about its equilibria (local Maxwellian) and the time derivative, i.e., $f_\alpha = f_\alpha^{(0)}+\epsilon f_\alpha^{(1)}+\epsilon^2f_\alpha^{(2)}+\cdots$ and $\partial_t = \partial_{t_0}+\epsilon \partial_{t_1}+\epsilon^2\partial_{t_2}+\cdots$, respectively, where $\epsilon$ is a small perturbation parameter. This is equivalent to the following expansions on the higher, non-conserved, raw moments
\begin{equation}
\tensr{\Gamma} = \tensr{\Gamma}^{(0)}+\epsilon \tensr{\Gamma}^{(1)}+\epsilon^2\tensr{\Gamma}^{(2)}+\cdots, \quad \tensr{\Lambda} = \tensr{\Lambda}^{(0)}+\epsilon \tensr{\Lambda}^{(1)}+\epsilon^2\tensr{\Lambda}^{(2)}+\cdots, \label{eq:CE_expansion_moments}
\end{equation}
in the above moment system. To the leading order, the mass and momentum equations in Eq.~(\ref{eq:zeroth_first_moment}) become
\begin{equation}
\partial_{t_0} \rho + \bm{\nabla}\cdot(\rho \bm{u}) =0, \quad \partial_{t_0} (\rho\bm{u}) + \bm{\nabla}\cdot \tensr{\Gamma}^{(0)} =\bm{F}, \label{eq:zeroth_first_moment_CE_expansion_leading_order}
\end{equation}
where $\tensr{\Gamma}^{(0)}=c_s^2\rho \tensr{I}+\rho\bm{u}\bm{u}$ is the equilibrium part of the second raw moment. On the other hand, the leading order
second raw moment equation, via Eq.~(\ref{eq:second_moment}), reads as
\begin{equation}
\partial_{t_0} \tensr{\Gamma}^{(0)} + \bm{\nabla}\cdot \tensr{\Lambda}^{(0)} =-\frac{1}{\tau} \tensr{\Gamma}^{(1)}+\tensr{\Upsilon}. \label{eq:second_moment_CE_expansion_leading_order}
\end{equation}
In order to recover the physically correct viscous stress, the non-equilibrium part of the second moment $\tensr{\Gamma}^{(1)}$ in the above equation, Eq.~(\ref{eq:second_moment_CE_expansion_leading_order}), should only be related to $\bm{\nabla}\cdot \tensr{\Lambda}^{(0)}$, which depends on the velocity gradients. However, the presence of the time derivative term in Eq.~(\ref{eq:second_moment_CE_expansion_leading_order}), i.e., $\partial_{t_0} \tensr{\Gamma}^{(0)}=c_s^2\partial_{t_0}\rho \tensr{I}+\partial_{t_0}(\rho\bm{u}\bm{u})$, in which the time derivatives of the velocity $\partial_{t_0}(\rho\bm{u}\bm{u})$ via the leading momentum equation (Eq.~(\ref{eq:zeroth_first_moment_CE_expansion_leading_order})) give rise to an additional term of the form $\bm{F}\bm{u}+\bm{u}\bm{F}$. This can be eliminated only if the corresponding moment of the body force $\tensr{\Upsilon}$ becomes equal to
\begin{equation}
\tensr{\Upsilon}=\bm{F}\bm{u}+\bm{u}\bm{F}. \label{eq:second_raw_moment_body_force}
\end{equation}
This necessary condition for the second raw moment of the body force $\sum_\alpha S_\alpha e_{\alpha i} e_{\alpha j}=F_iu_j+F_ju_i$, which is a classic result of the acceleration term in the Boltzmann equation, was given in~\cite{luo1998unified}. This implies a vanishing second central moment of the body force, i.e., $\sum_\alpha S_\alpha (e_{\alpha x}-u_x)^m (e_{\alpha y}-u_y)^n=0$ for $m+n=2$, which appears explicitly in~\cite{wagner2006thermodynamic} and was considered in the previous unsplit forcing approach for the cascaded LB scheme~\cite{Premnath2009for}.

In view of the above, in our present operator-split forcing based cascaded LB formulation, the PDE needed for the solving the split force step given in Eq.~(\ref{eq:13}) is a central moment representation of the split kinetic equation $\partial_t  f_{\alpha}=S_{\alpha}$. That is, taking the central moments of this equation of order $(m+n)$, we get an evolution equation as follows:
\begin{equation}
\textbf{Step}\,  \tensr F: \frac{\partial}{\partial t}\widehat{\kappa}_{x^my^n}=\widehat{\sigma}_{x^my^n},
\end{equation}
where $\widehat{\kappa}_{x^my^n}=\sum_\alpha f_\alpha (e_{\alpha x}-u_x)^m (e_{\alpha y}-u_y)^n$ and $\widehat{\sigma}_{x^my^n}=\sum_\alpha S_\alpha (e_{\alpha x}-u_x)^m (e_{\alpha y}-u_y)^n$ are the central moments of the distribution functions and the source term due to the body force, respectively. It thus follows that, in particular, the continuous time equations for the change in the second central moment components for the split body force step are given as
\begin{equation}
\textbf{Step}\,  \tensr F: \frac{\partial}{\partial t}\widehat{\kappa}_{xx}=0, \quad \frac{\partial}{\partial t}\widehat{\kappa}_{yy}=0, \quad \frac{\partial}{\partial t}\widehat{\kappa}_{xy}=0,
\end{equation}
which implies the necessary condition for introducing the canceling second raw moment components of the body force, i.e., $2F_xu_x$, $2F_yu_y$ and $F_xu_y+F_yu_x$ to eliminate the spurious effects in the viscous stress and thereby correctly recover the Navier-Stokes equations as mentioned above.

\section{\label{app:5}Extension of the Symmetrized Operator Split Implementation for Cascaded LB Method for Passive Scalar Transport Including Sources}
In many applications, the transport of a passive scalar (e.g., temperature or species concentration) occurs, which is generally represented by means of the following convection-diffusion equation (CDE) with a source term
\begin{equation}
\partial_t\phi+\bm \nabla \cdot (\bm u \phi)=\bm \nabla \cdot(D_{\phi}\bm \nabla \phi)+S_{\phi}.
\label{eq:33}
\end{equation}
Here, $\phi$ is the passive scalar variable, $D_{\phi}$ is the diffusion coefficient, and $S_{\phi}$ is the local source term (e.g. due to viscous dissipation, internal heat generation or chemical reaction). Various LB schemes have been investigated for modeling the CDE during the last two decades (e.g.,~\cite{ponce1993lattice,he1998novel,van2000convection,lallemand2003theory,rasin2005multi,chopard2009lattice,yoshida2010multiple,wang2013lattice,Chai2013,Contrino2014}). A novel numerical approach considered in this study for the solution of Eq.~(\ref{eq:33}) is as follows. The velocity $\bm u$ in the above equation can be obtained from the cascaded LB scheme for the D2Q9 lattice presented in the previous section. Our goal is to solve for the passive scalar field $\phi$ whose evolution is represented by the above CDE, but without the source term using a separate 2D cascaded scheme with collide and stream steps involving another distribution function; then implement the effect of the source term $S_{\phi}$ via additional source steps using an operator split scheme based on a symmetrization principle. To meet this objective, we consider a new cascaded LB scheme for coupled fluid flow and scalar transport that we developed recently in different dimensions~\cite{Hajabdollahi2018} and further accelerated by using multigrid~\cite{Hajabdollahi2017b}. Here, a two-dimensional, five velocity (D2Q5) lattice based cascaded LB method is introduced to represent the evolution of the passive scalar field via the CDE, which is adopted in this work for further extension using an operator split source implementation.

The D2Q5 lattice is represented by means of the following components of the particle velocity vectors $\ket{e_{x}}$ and $\ket{e_{y}}$:
\begin{subequations}
\begin{eqnarray}
&\ket{e_{x}} =\left(     0,     1,    0,     -1,     0  \right)^\dag,\label{eq:34a}\\
&\ket{e_{y}} =\left(     0,     0,     1,     0,    -1\right)^\dag\label{eq:34b}.
\end{eqnarray}
\end{subequations}
In addition, we introduce the following $\ket{1}$ vector:
\begin{equation}
\ket{1} =\left(     1,     1,     1,     1,    1\right)^\dag.\label{eq:34c}
\end{equation}
The zeroth moment is the Euclidean inner product of this vector with the distribution functions.
The corresponding five orthogonal basis vectors are given by~\cite{Hajabdollahi2017b}
\begin{eqnarray}
\ket{L_0}=\ket{1},\,
\ket{L_1}=\ket{e_{x}},\,
\ket{L_2}=\ket{e_{y}},\nonumber \\
\ket{L_3}=5\ket{e_{x}^2+e_{y}^2}-4\ket{1},\,
\ket{L_4}=\ket{e_{x}^2-e_{y}^2},\label{eq:35}
\end{eqnarray}
which can be grouped together as the following transformation matrix $\tensr{L}$ for mapping changes in the moment space to those in the velocity space
\begin{equation}
\tensr{L}=\left[\ket{L_0},\ket{L_1},\ket{L_2},\ket{L_3},\ket{L_4}\right].
\label{eq:36}
\end{equation}
In order to represent the structure of the cascaded collision operator for the passive scalar field, we define the following central moments and raw moments, respectively, of the distribution function $g_{\alpha}$ and its equilibrium $g_{\alpha}^{eq}$ as
\begin{eqnarray}
\left( \begin{array}{l}
{{\hat \kappa }_{{x^m}{y^n}}}^{\phi}\\
\hat \kappa _{{x^m}{y^n}}^{eq,\phi}
\end{array} \right) = \sum\limits_\alpha  {\left( \begin{array}{l}
{g_\alpha }\\
g_\alpha ^{eq}\\
\end{array} \right)} {({e_{\alpha x}} - {u_x})^m}{({e_{\alpha y}} - {u_y})^n},\label{eq:37}
\end{eqnarray}
and
\begin{eqnarray}
\left( {\begin{array}{*{20}{l}}
{{{\hat \kappa }_{{x^m}{y^n}}}}^{\phi'}\\
{\hat {\kappa} _{{x^m}{y^n}}^{{eq,\phi'}}}
\end{array}} \right) = \sum\limits_\alpha  {\left( {\begin{array}{*{20}{l}}
{{g_\alpha }}\\
{g_\alpha ^{eq}}
\end{array}} \right)} {e_{\alpha x}^m}{e_{\alpha y}^n}.
\label{eq:38}
\end{eqnarray}
By equating the discrete central moments of the equilibrium distribution function with the corresponding continuous central moments based on the local Maxwellian (wherein the density is replaced by $\phi$), we get
\begin{eqnarray}
\widehat{\kappa}^{eq,\phi}_{0}=\phi,\,
\widehat{\kappa}^{eq,\phi}_{x}=0,\,
\widehat{\kappa}^{eq,\phi}_{y}=0,\,
\widehat{\kappa}^{eq,\phi}_{xx}=c_{s\phi}^2\phi,\,
\widehat{\kappa}^{eq,\phi}_{yy}=c_{s\phi}^2\phi, \label{eq:39}
\end{eqnarray}
which will be used in the construction of the collision operator later. In this work, wet set $c_{s\phi}^2=1/3$. Then, the 2D cascaded LB scheme for the passive scalar transport without the source term can be represented by means of the following collision and streaming steps:
\begin{subequations}
\begin{eqnarray}
 & g^{p}_{\alpha}=g_{\alpha}+(\tensr L\cdot \widehat {\mathbf h})_{\alpha},
\label{eq:40a}\\
& g_{\alpha}(\bm x, t)=g^{p}_{\alpha}(\bm x-\bm e_{\alpha}\Delta t,t).
\label{eq:40b}
\end{eqnarray}
\end{subequations}
The procedure to obtain the change of different moments under cascaded collision, i.e. $\widehat {\mathbf h}$ based on the central moment equilibria Eq.~(\ref{eq:39}) is analogous to that used in the previous section for fluid flow, with the main difference
being that in the present case, there is only one collisional invariant, i.e. $\phi$, and hence $\widehat { h}_0=0$. Then, it follows that~\cite{Hajabdollahi2017b} (see also~\cite{Hajabdollahi2018} that elaborates such a formulation for a 3D cascaded LBM for CDE)
\begin{eqnarray}
\widehat{h}_1&=&\frac{\omega_1^\phi}{2}\left[\phi u_x-{\widehat\kappa}_x^{\phi'}\right],  \nonumber \\
\widehat{h}_2&=&\frac{\omega_2^\phi}{2}\left[\phi u_y-{\widehat\kappa}_y^{\phi'}\right], \nonumber \\
\widehat{h}_3&=&\frac{\omega_3^\phi}{4}\left[2c_{s\phi}^2\phi-({{\widehat\kappa}_{xx}}^{\phi'}+{{\widehat\kappa}_{yy}}^{\phi'})+2(u_x{\widehat\kappa}_x^{\phi'}+u_y{\widehat\kappa}_y^{\phi'})+(u_x^2+u_y^2)\phi\right]+u_x\widehat{h}_1+u_y\widehat{h}_2, \nonumber \\
\widehat{h}_4&=&\frac{\omega_4^\phi}{4}\left[-({{\widehat\kappa}_{xx}}^{\phi'}-{{\widehat\kappa}_{yy}}^{'\phi})+2(u_x{\widehat\kappa}_x^{\phi'}-u_y{\widehat\kappa}_y^{\phi'})+(u_x^2-u_y^2)\phi\right]+u_x\widehat{h}_1-u_y\widehat{h}_2.
\label{eq:41}
\end{eqnarray}
where $\omega_1^\phi$, $\omega_2^\phi$, $\omega_3^\phi$ and $\omega_4^\phi$ are the relaxation parameters. Notice that the cascaded structure of the expressions for the change of moments $\widehat {\mathbf h}$ starts from the first order moments for the CDE, unlike those for the NSE given the previous section. The relaxation parameters for the first order moments in the above determine the molecular diffusivity $D_{\phi}$: $D_{\phi}=c_{s\phi}^2(\frac{1}{\omega_j^\phi}-\frac{1}{2})\Delta t$, $j=1,2$. After the streaming step in Eq.~(\ref{eq:40b}), the output passive scalar field $\phi^o$ is obtained as the zeroth moment of $g_{\alpha}$ as
\begin{equation}
\phi^o=\sum_{\alpha=0}^4 g_{\alpha}.
\label{eq:42}
\end{equation}
The effect of the source term $S_{\phi}$ can then be introduced as the solution of the source subproblem split from Eq.~(\ref{eq:33}): $\partial_t{\phi}=S_{\phi}$. As before, this can be implemented by means of two symmetrized sequence of steps before and after collision, each using a time step $\Delta t/2$ and such a source operator will be denoted by $\tensr R^{1/2}$. Thus, the extension of the Strang splitting approach for the cascaded LBM to represent the source term in the CDE can be formulated as
\begin{eqnarray}
 {g_{\alpha}}(\bm x,t+\Delta t)=\tensr S\, \tensr R^{1/2}\,\tensr C\, \tensr R^{1/2}{g_{\alpha}}(\bm x,t).
\label{eq:StrangLBMCDE}
\end{eqnarray}

Solving the above subproblem representing the evolution of the scalar field $\phi$ due to the source term $S_{\phi}$ yields the following step before collision
\begin{eqnarray}
\text{Pre-collision Source Step}\,\tensr R^{1/2}: \phi= \phi^o+ {\frac {S_\phi}{2}}\Delta t.
\label{eq:43}
\end{eqnarray}
This updated $\phi$ is then used to perform the cascaded collision relaxation step and determine the change of different moments under collision $\widehat {h}_{\beta}$, where $\beta=1,2,3,4,$ given in Eq.~(\ref{eq:41}). Analogously, the other source half step following collision can be represented as
\begin{equation}
\text{Post-collision Source Step}\,\tensr R^{1/2}: \phi^p= \phi+ {\frac {S_\phi}{2}}\Delta t=\phi^o+S_{\phi}\Delta t.
\label{eq:44}
\end{equation}
In order to effectively implement this in the cascaded formulation, we take the zeroth moment of the post-collision distribution $g_{\alpha}^p$ given by $g_{\alpha}^p={g_\alpha}+(\tensr L\cdot \widehat {\mathbf h})_{\alpha}$, which yields
\begin{equation}
 \phi^p=\sum_{\alpha}{g^p_{\alpha}}=\sum_{\alpha}{g_{\alpha}}+\sum_{\beta}{\braket{{K_{\beta}}|1}} \widehat h_{\beta}.
\label{eq:45}
\end{equation}
Based on the orthogonal basis vectors given in Eq.~(\ref{eq:35}), it follows that $\sum_{\beta}{\braket{{K_{\beta}}|1}} \widehat h_{\beta}=5\widehat h_0$, which when substituted in Eq.~(\ref{eq:45}), and along with Eq.~(\ref{eq:42}), we obtain
\begin{equation}
 \phi^p=\phi^o+5\widehat{h}_0.
\label{eq:46}
\end{equation}
Comparing the target result Eq.~(\ref{eq:44}) with the above constructed field (Eq.~(\ref{eq:46})), we get the following result for the zeroth order moment change due to the source $S_\phi$
\begin{equation}
 \widehat{h}_0=\frac{S_{\phi}}{5}\Delta t.
\label{eq:47}
\end{equation}
This effectively implements the effect of the post-collision source step over a step length of $\Delta t/2$. Using this result (Eq.~(\ref{eq:47})) along with Eq.~(\ref{eq:41}) for the change of moments under collision in Eq.~(\ref{eq:40a}) and expanding $(\tensr K\cdot \widehat{\tensr h})_{\alpha}$, we obtain the post-collision distribution functions, which read as
\begin{eqnarray}
{g}_{0}^p&=&{g}_{0}+\left[\widehat{h}_0-4\widehat{h}_3\right],\nonumber \\
 {g}_{1}^p&=&{g}_{1}+\left[\widehat{h}_0+\widehat{h}_1+\widehat{h}_3+\widehat{h}_4\right],\nonumber \\
{g}_{2}^p&=&{g}_{2}+\left[\widehat{h}_0+\widehat{h}_2+\widehat{h}_3-\widehat{h}_4\right], \nonumber  \\
{g}_{3}^p&=&{g}_{3}+\left[\widehat{h}_0-\widehat{h}_1+\widehat{h}_3+\widehat{h}_4\right],\nonumber \\
 {g}_{4}^p&=&{g}_{4}+\left[\widehat{h}_0-\widehat{h}_2+\widehat{h}_3-\widehat{h}_4\right].
 \label{eq:48}
\end{eqnarray}

The overall sequence of computational steps for the 2D cascaded LB scheme for passive scalar transport with a source implementation based on the Strang splitting is similar to that for the fluid flow presented in the previous section. Moreover, such a symmetrized operator splitting formulation can
also be used to represent forces/sources in the 3D central moment based LBM for thermal convective flows developed recently~\cite{Hajabdollahi2018}.

\section{\label{Results}Results and Discussion}
We will now present a numerical validation study of the various symmetrized operator split schemes to incorporate forces/sources in the cascaded LB method presented earlier by comparison of  their computed results against a set of benchmark problems with analytical solutions. In the following, all the numerical results will be generally reported in the lattice units typical for LB simulations~\cite{Kruger2016}. That is, unless otherwise specified, we consider $\Delta x = \Delta t=1$ and hence the particle speed $c=\Delta x/\Delta t$ is taken to be unity. The fluid velocity will be scaled by the particle speed $c$, and the reference scale for the density $\rho_0$ is 1.0. For the cascaded LB method for fluid flow presented in Sec.~\ref{app:d3q27matrix}, the considerations for the relaxation parameters are as follows: $\omega_4$ and $\omega_5$ determine the shear kinematic viscosity (via $\omega_4=\omega_5=1/\tau$ and $\nu=\frac{1}{3}(\tau-\frac{1}{2})\Delta t$), which can be specified from the problem statement. The parameter $\omega_3$ is related to the bulk viscosity (see e.g.~\cite{Premnath2009for}), while the remaining parameters for the higher order moments $\omega_6, \omega_7$ and $\omega_8$, along with $\omega_3$ can be tuned to improve numerical stability. A detailed study of the influence of such parameters in the cascaded LB scheme was performed in~\cite{ning2016numerical}. For turbulent flow computations, care needs to be exercised in choosing
the relaxation parameters for the higher order moments in order to avoid being over-dissipative. In this work, for the incompressible, laminar flow benchmark flow problems considered in the following, we use $\omega_3=\omega_6=\omega_7=\omega_8=1.0$. On the other hand, for the cascaded LB method for the solution of the passive scalar transport presented in Sec.~\ref{app:5}, the parameters $\omega_1^\phi$ and $\omega_2^\phi$, which are related to the coefficient of diffusivity (i.e. $\omega_1^\phi=\omega_2^\phi=1/\tau^\phi$ and $D_\phi=\frac{1}{3}(\tau^\phi-\frac{1}{2})\Delta t$), are assigned from the problem statement based on the characteristic dimensionless group; relaxation parameters $\omega_j^\phi$, where $j=3,4,5$, which influence the numerical stability, are set to unity in this work.

\subsection{Poiseuille Flow}
\begin{figure}[htbp]
\centering
\advance\leftskip-1cm
\advance\rightskip-1cm
      {
        \includegraphics[ height=7cm, width=7cm] {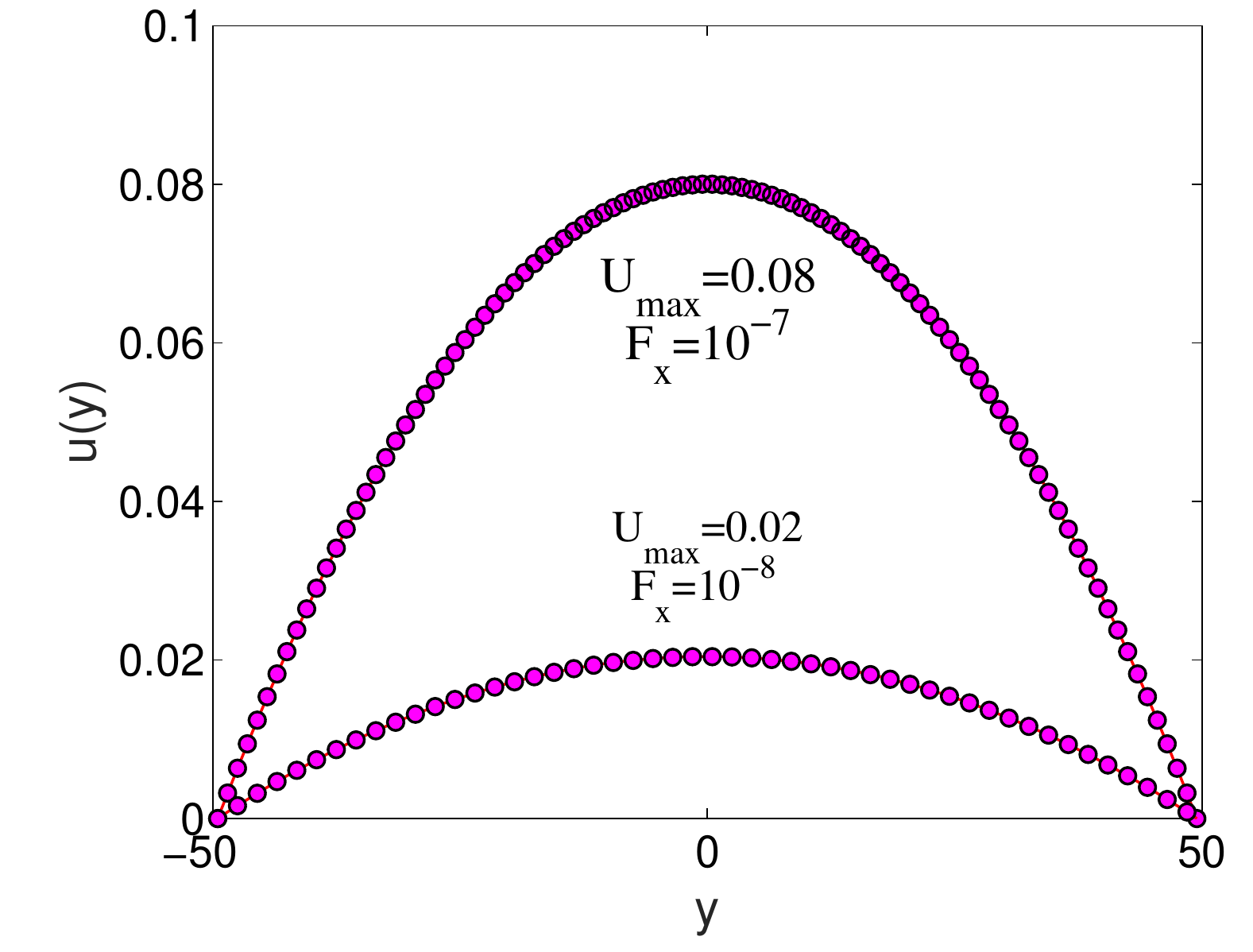}
        }
        \caption{Comparison of the computed velocity profiles using the 2D symmetrized operator split cascaded LB forcing scheme with the analytical solution for Poiseuille flow for body force magnitudes of $10^{-7}$ and $10^{-8}$. The lines indicate the analytical results, and the symbols are the solutions obtained by our present numerical scheme.}
    \label{fig:poi}
\end{figure}

In these sections, we validate our 2D operator split forcing approach by considering various test problems involving different types of body force fields. For the first problem, a two-dimensional Poiseuille flow in a channel discretized with $3\times 100$  lattice nodes is considered. In our computations, at the top and bottom walls, a no-slip boundary condition, and at the inlet and outlet, periodic boundary conditions are applied. The no-slip boundary condition is implemented by using the classical half-way bounce back scheme in this work~\cite{ladd1994numerical,Kruger2016}. The analytical solution of the velocity profile flow for this laminar flow problem can be written as follows: $u(y)=U_{max}[1-(\frac{y}{L})^2]$, where $U_{max}=F_xL^2/(2\rho \nu)$ is the maximum velocity along the central line. Here, $L$, $\rho$ and $\nu$ are the channel half-width, fluid density and kinematic viscosity respectively. $F_x$ is a constant body force acting in the $x$-direction which drives the flow. Comparison of the  simulation results of the velocity profile against the analytical solution is shown in Fig.~\ref{fig:poi}, where the body forces for two cases with maximum velocities of $0.02$ and $0.08$ are set to the values of $10^{-8}$ and $10^{-7}$, respectively. For the former case, the relaxation time $\tau$ is chosen to be $0.5019$, which for the latter it is $0.5047$. The corresponding Mach numbers $\mbox{Ma}$ are $0.034$ and $0.138$, respectively. It can be clearly seen that there is an excellent agrement between the numerical simulation carried out using the 2D symmetrized operator split cascaded LB forcing scheme and the analytical solution for the both cases.

\subsubsection*{Grid Convergence Study}
\begin{figure}[htbp]
\centering
\advance\leftskip-1cm
\advance\rightskip-1cm
      {
        \includegraphics[scale=0.5] {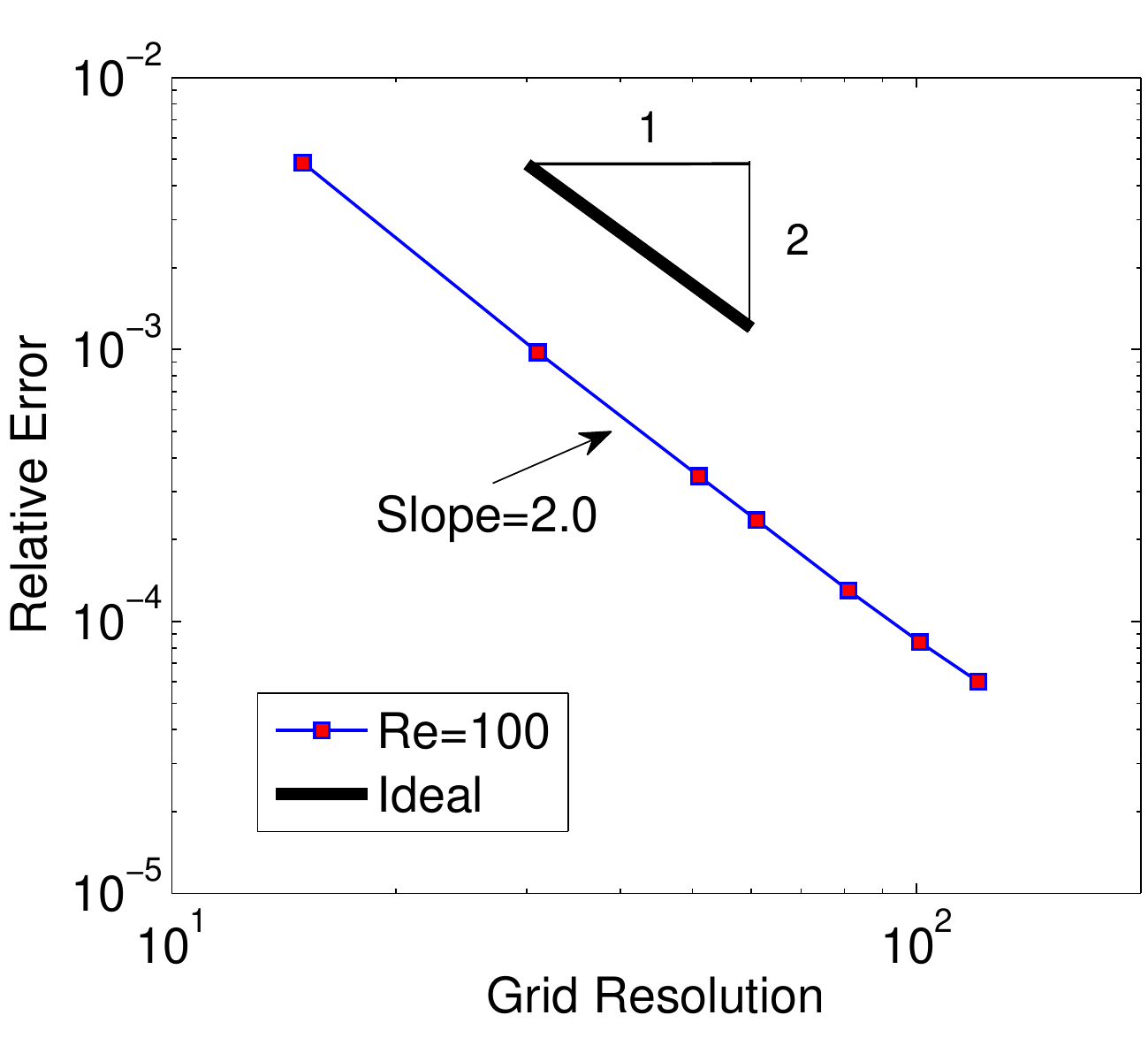}
        \label{fig:img2} }
        \caption{Grid convergence for 2D Poiseuille flow with a constant Reynolds number $Re=100$ and relaxation time $\tau=0.55$ computed using the 2D symmetrized operator cascaded LB forcing scheme.}
        \label{fig:sca}
\end{figure}

In order to determine the order of accuracy of our symmetrized operator split forcing scheme, we perform a grid convergence test by applying a diffusive scaling. According to this scaling, Mach number $\mbox{Ma}=U/c_s$ reduces proportionally with the increase in the grid resolution at a fixed viscosity or fixed relaxation time $\tau=1/\omega_j$, $j=4,5$, where $\omega_4$ and $\omega_5$ represent the relaxation parameters for the second order moments in the 2D cascaded LB scheme (see Sec. 4), so that the scheme has asymptotic convergence to the incompressible flow limit. For our simulation, we consider a Poiseuille flow with the same set up as considered earlier. We consider a sequence of $3\times 15,3\times 31,\dots, 3\times 121$ lattice nodes to study grid convergence under diffusive scaling when the relaxation time and Reynolds number are set to $\tau=0.55$ and $100$, respectively. Next, to quantify the grid convergence, we consider the global relative error ($E_{g,u}$) of the flow field under a discrete $\ell_2$-norm as follows:
\begin{eqnarray}
\|E_{g,u}\|_2=\sqrt{\frac{\Sigma (u_c-u_a)^2}{\Sigma{(u_a)^2}}},
\label{eq:sca}
\end{eqnarray}
where $u_c$ and $u_a$ is the computed and analytical solutions, respectively, and the summation is carried out for the flow domain. The relative error between the computed results and the analytical solution against different grid resolutions is illustrated in Fig.~\ref{fig:sca}. The relative errors have a slope of $2.00$ which indicates that our new approach based on the symmetrized operator split forcing scheme for the cascaded LB method is spatially second-order accurate.

\subsection{Hartmann Flow}
\begin{figure}[htbp]
\centering
\advance\leftskip-1cm
\advance\rightskip-1cm
      {
        \includegraphics[ height=7cm, width=7cm] {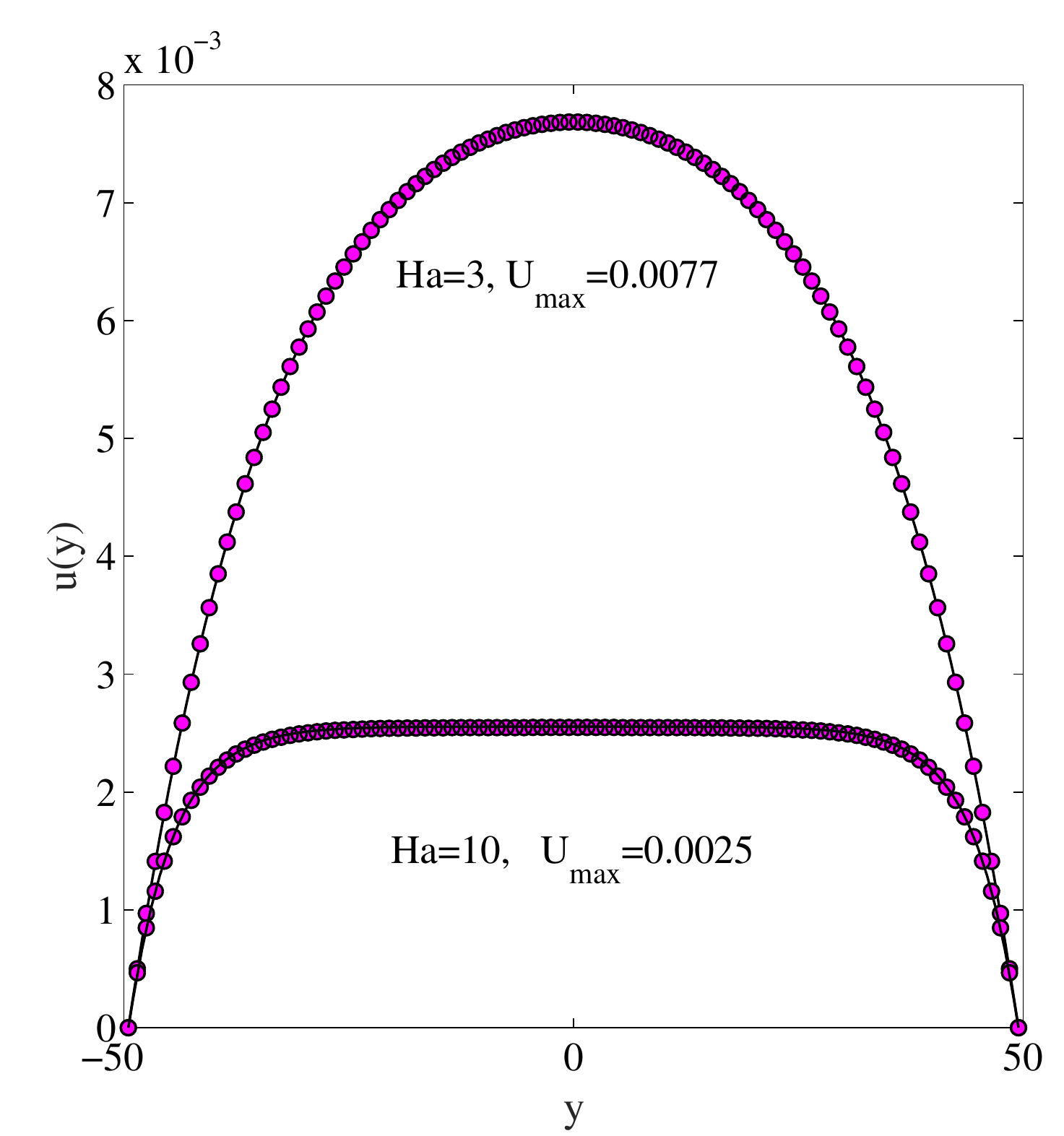}
        \label{fig:img2} }
        \caption{Comparison of the computed velocity profiles using the 2D symmetrized operator split cascaded LB forcing scheme with the analytical solution for Hartmann flow for Hartmann numbers $\mbox{Ha}$ of 3 and 10. The lines indicate the analytical results, and the symbols are the solutions obtained by our present numerical scheme.}
    \label{fig:har}
\end{figure}

As the next benchmark case study, a numerical comparison of the results with our 2D operator split forcing approach is made for a specific type of magnetohydrodynamic (MHD) flow, i.e., the flow between two unbounded plates subjected to a transverse magnetic field known as the Hartmann flow. This type of flow arises in a variety of engineering devices including MHD pumps, fusion devices, generators and microfluidic devices. Furthermore, an inherent spatially-varying body force makes this benchmark a particularly suitable test problem for the present study. The fluid is driven by a constant body force $F_b$ and retarded  by a local variable force (i.e. Lorentz force) arising by an interaction between a uniform steady magnetic field $B_y=B_0$, acting perpendicular to the channel walls and the fluid motion. By choosing the $x$ axis for the  flow direction and the $y$ axis to be co-directional with the external magnetic field $B_y=B_0$, the induced magnetic field resulting from such an interaction can be represented as $B_x(y)=\frac{F_bL}{B_0}\left[\frac{\mbox{sinh}\left(\mathrm{Ha}\frac{y}{L}\right)}{\mbox{sinh}(\mathrm{Ha})}-\frac{y}{L}\right]$. Here, the Hartmann number, $\mbox Ha$ is the square root of the ratio of the electromagnetic force to the viscous force and  $L$ and $F_b$ are the channel half-width and the uniform driving force, respectively. Consequently, the effectively spatially varying body force which act on the flow is $F_x=F_b+F_{mx}$ . This is a combination of the Lorentz force $F_{mx}=B_0\frac{dB_x}{dy}$, and the uniform driving force $F_b$. The analytical solution for such a problem is $u_x(y)=\frac{F_bL}{B_0}\sqrt{\frac{\eta}{\nu}}\mbox{coth}(\mathrm{Ha})\left[1-\frac{\mbox{cosh}\left(\mathrm{Ha}\frac{y}{L}\right)}{\mbox{cosh}(\mathrm{Ha})}\right]$. Here, $\nu$ is the kinematic viscosity and $\eta$ is the magnetic resistivity, which can be represented by $\eta={B_0}^2L^2/{\mbox Ha^2\nu}$. We consider the same set up as considered for the Poiseuille flow simulation for the boundary conditions but now with spatially varying body forces. For two different values of $\mbox Ha$, 3 and $10$, corresponding to Mach numbers of 0.013 and 0.004, respectively, the computed velocity profiles against the analytical solution are illustrated in Fig.~\ref{fig:har}. It can be observed that the present simulation is able to reproduce the analytical solution very well. In particular, the significant flattening of the velocity profile at higher $\mbox Ha$ is well reproduced by our forcing scheme.

\subsection{Womersley Flow}

\begin{figure}[htbp]
\centering
\advance\leftskip.5cm
\advance\rightskip1.5cm
    \subfloat{
        \includegraphics[ height=6cm, width=6cm] {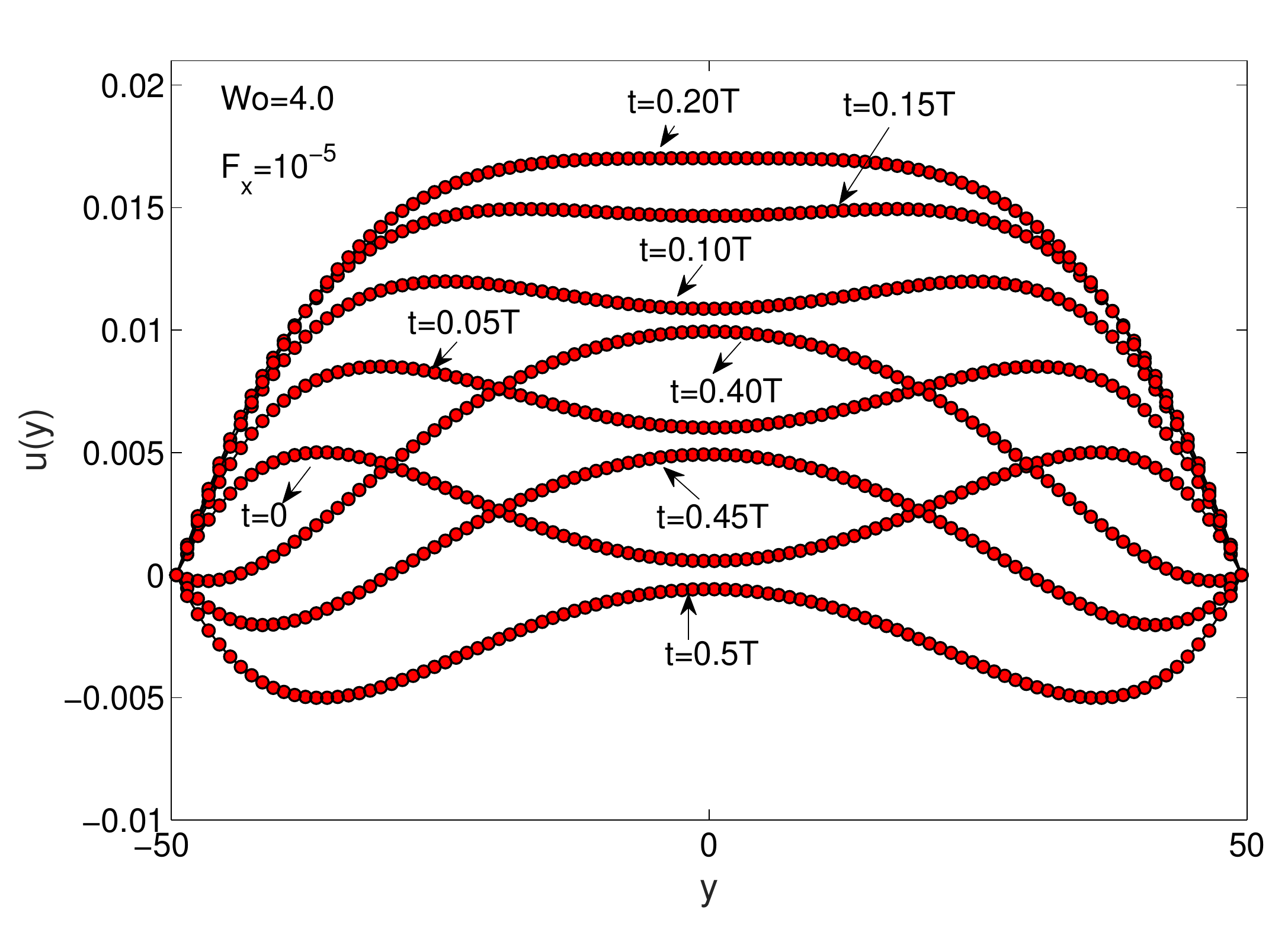}
        \label{fig:img12} } \hspace*{-22em}
    \hfill
    \subfloat{
        \includegraphics[ height=6cm, width=6cm] {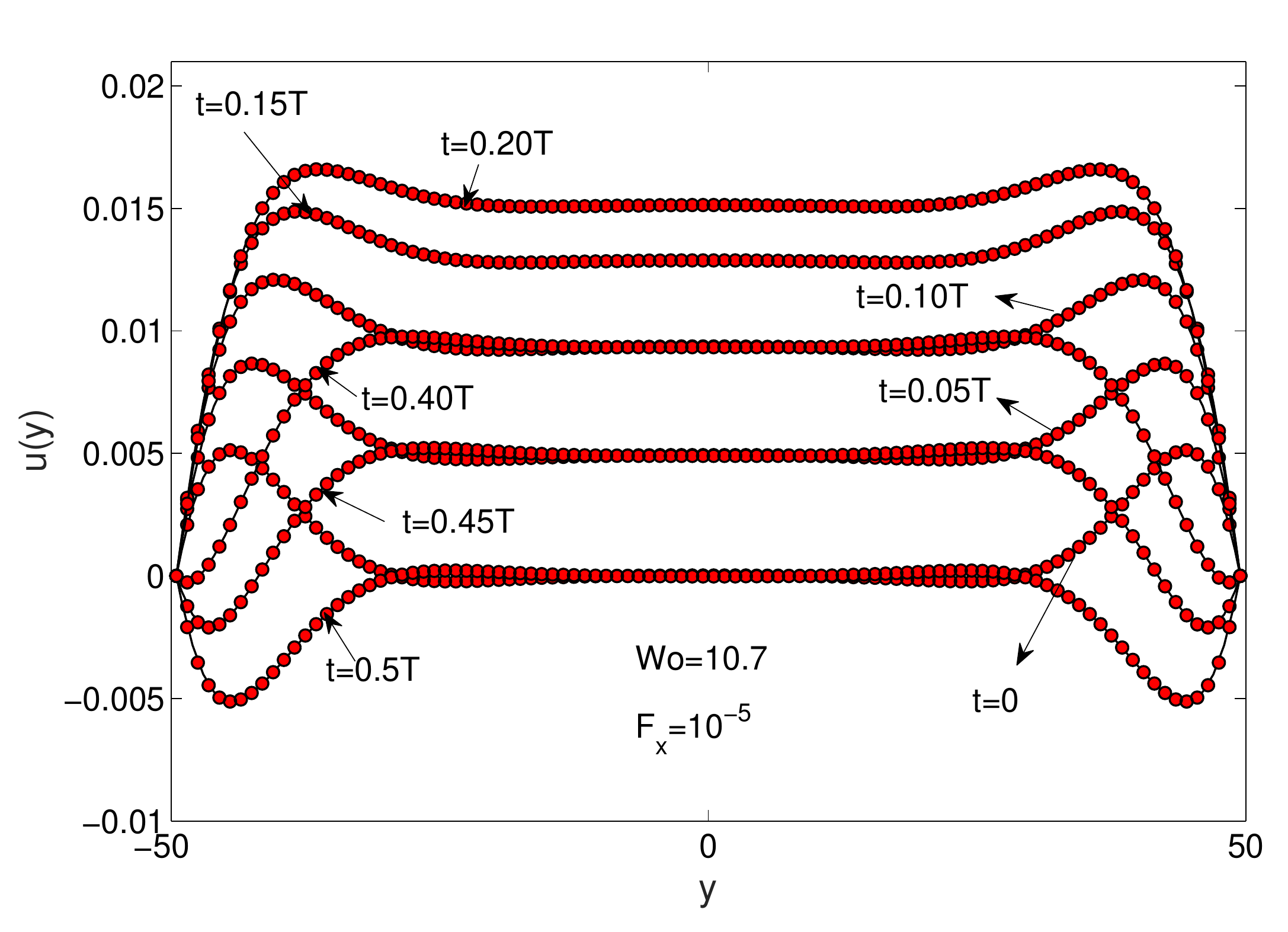}
        \label{fig:img24} } \\
    \caption{Comparison of computed and analytical velocity profiles at different instants within a time period of  pulsatile flow at two different Womersley numbers of  \mbox{Wo = 4} \mbox{Wo = 10.7}. Here, lines represent the analytical solution and symbols refer to the numerical results obtained using the 2D symmetrized operator split cascaded LB forcing scheme.}
    \label{fig:wom}
\end{figure}

We now turn to study the Womersley flow, which is a flow between two infinite parallel plates driven by a temporally oscillatory external force. This benchmark problem is used to assess the ability of our symmetrized operator split forcing scheme for  representing time-dependent body forces. The external force $F_m\mbox{cos}(\omega t)$ oscillates with an amplitude $F_m$ and with an angular frequency $\omega=2 \pi/T$, where $T$ is the time period. Supposing that the flow is laminar and incompressible, the analytical solution for the velocity field is given as
\begin{eqnarray}
&u(y,t)=\mbox{Re}\left\{i\frac{F_m}{\omega}\right[1-\frac{\mbox {cos}(\gamma y/L)}{\mbox {cos}{\gamma}}]e^{(i\omega t)} \},
\label{eq:wo}
\end{eqnarray}
where $\gamma=\sqrt{i\mbox{Wo}^2}$, $\mbox{Wo}= L \sqrt{(\omega/\nu)}$ being the Womersley number, which is a non-dimensional parameter representing  the ratio of the channel half width $L$ to the diffusion length over an oscillation period (i.e., the Stokes layer thickness). $\mbox{Re}\left\{\cdot\right\}$ represents taking the real part of the expression within the brackets. The simulation parameters are set as follows. The computational domain is resolved by a $3\times 100$  mesh, the time period $T=10000$ and the maximum force amplitude is set to $F_m=1\times10^{-5}$. The boundary condition at the inlet and the outlet is periodic and the half-way bounce-back scheme to represent the no-slip condition is used at the walls. The body force for this case is implemented as a solution of Eq.~(\ref{eq:13}) to update the velocity field. Since the explicit form of the time-dependent force is known here, it can be either analytically integrated to perform the velocity update in the force step or solved numerically by representing the body force $F_x$ via the trapezoidal rule as $\frac{1}{2}F_m(\cos(\omega t)+ \cos(\omega t+\Delta t/2))$. The latter approach is used in the present study. In general cases, if the body force $\bm{F}$ depends on $\bm{u}$, then Eq.~(\ref{eq:13}) needs to be numerically integrated and used as an implicit equation to solve for $\bm{u}$. Simulations are carried out to obtain the velocity profiles across the channel at different time instants with the time period $T$. Figure~\ref{fig:wom} shows a comparison for the velocity profiles for two values of the Womersley number, i.e. $4$ and $10.7$ at different time instants. It can be clearly seen that the numerical results agree well with the analytical solution represented by Eq.~(\ref{eq:wo}). Thus, the symmetrized operator split forcing scheme is able to represent flow profiles driven by time varying body forces with excellent accuracy.

\subsection{\label{Results_duct}Flow through a Square Duct}
\begin{figure}[htbp]
\centering
\advance\leftskip-1cm
\advance\rightskip-1cm
      {
        \includegraphics[scale=0.5] {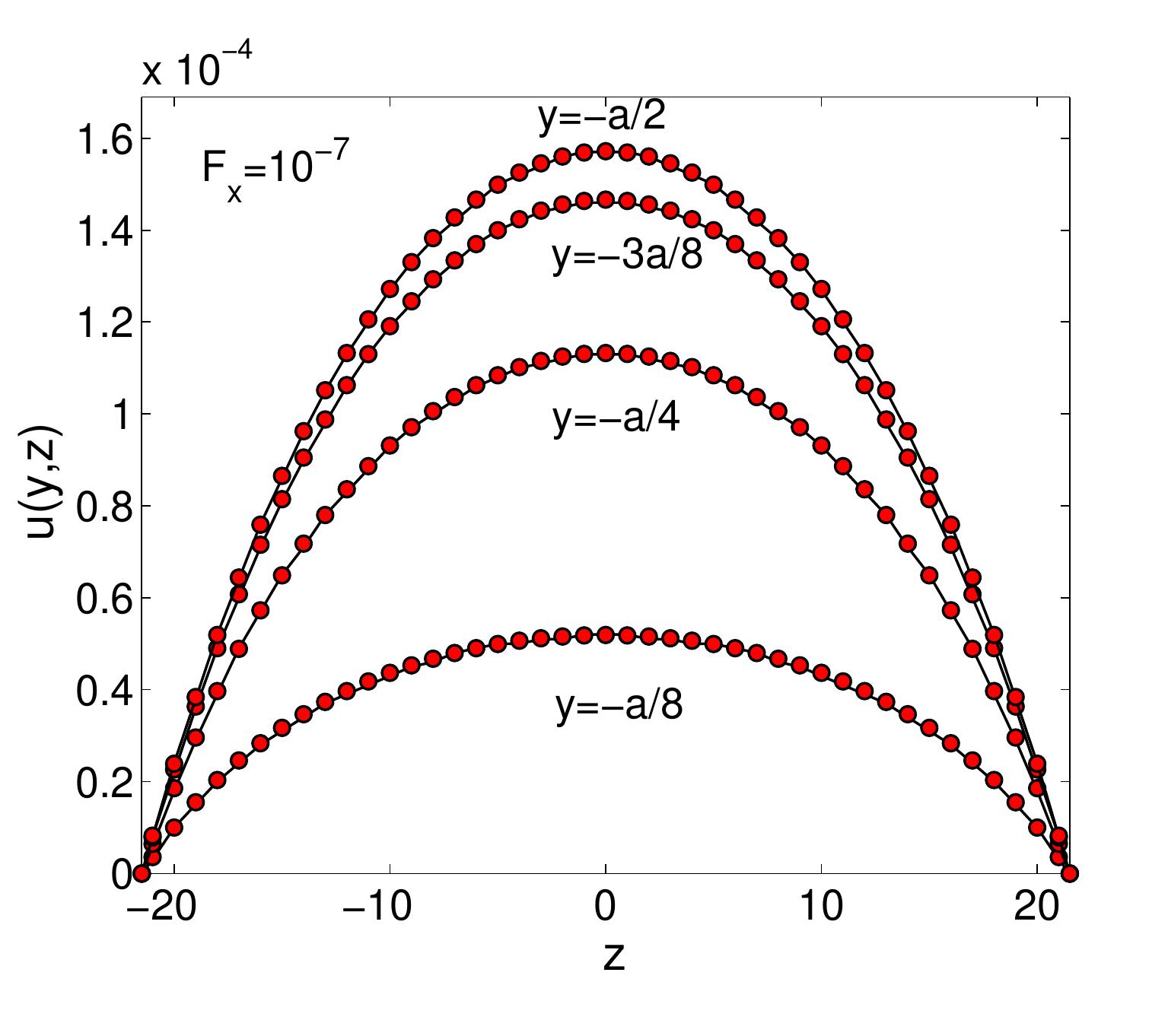}
        \label{fig:img2} }
        \caption{Comparison of the computed  velocity profiles using the 3D symmetrized operator split cascaded LB forcing scheme and the analytical solution, for flow through a square duct in presence of a body force magnitude of $F_x=10^{-7}$  for different values of $y$. Here, lines represent the analytical solution and symbols refer to the results obtained using the present numerical scheme.}
    \label{fig:duc}
\end{figure}

In order to validate our 3D symmetrized operator split forcing scheme for a multidimensional flow subjected to a body force, we consider flow through a square duct driven by a constant body force $F_x$. In our computations, we apply periodic boundary conditions at the inlet and outlet and a no-slip boundary condition at the four wall surfaces. For a channel with width $2a$, this test problem has an analytical solution based on a Fourier series for the velocity field, which reads as
\begin{eqnarray}
u(y,z) = \frac{{16\mathop a\nolimits^2 \mathop F\nolimits_x }}{{\rho \nu \mathop \pi \nolimits^3 }}\sum\limits_{n = 1}^\infty  {\mathop {( - 1)}\nolimits^{(n - 1)} \left[1 - \frac{{\cosh (\frac{{(2n - 1)\pi z}}{{2a}})}}{{\cosh (\frac{{(2n - 1)\pi }}{2})}}\right]} \frac{{\cos \left(\frac{{(2n - 1)\pi y}}{{2a}}\right)}}{{\mathop {(2n - 1)}\nolimits^3 }},
\end{eqnarray}
where $\rho$ and $\nu$ are the fluid density and kinematic viscosity, respectively and $x$ is the direction of the flow, and $-a<y<a$, $-a<z<a$ is the cross section of the duct. We chose a grid resolution of $3\times45\times45$, with a relaxation parameter $\tau$ equal to $0.76$, and a body force magnitude of $F_x=1\times 10^{-7}$ is applied. Figure~\ref{fig:duc} illustrates the velocity profiles $u(y,z)$ computed using our $3D$ symmetrized operator split scheme to incorporate forcing terms in the 3D cascaded LB method for different values of $y$. In this figure, a comparison with the analytical solution given above is also made. It is evident that there is a very good agreement between our computed results and the analytical solution for this body force driven multidimensional flow problem.

\subsection{Four-rolls Mill Flow Problem}
\begin{figure}[htbp]
\centering
\advance\leftskip-.8cm
\advance\rightskip .3cm
    \subfloat[Present work]{
        \includegraphics[scale=0.50] {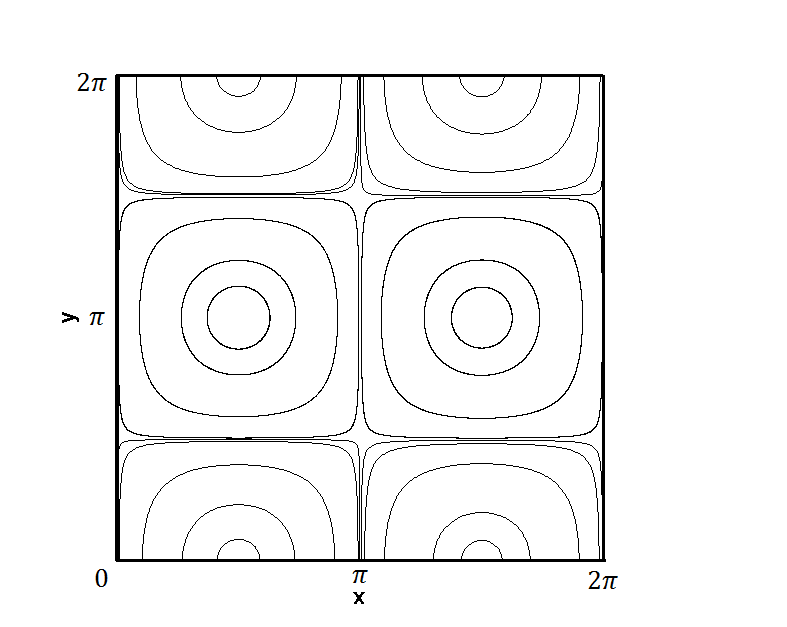}
        \label{fig:img1} } \hspace*{-5em}
    \hfill
   \subfloat[Analytical]{
         \includegraphics[scale=0.50] {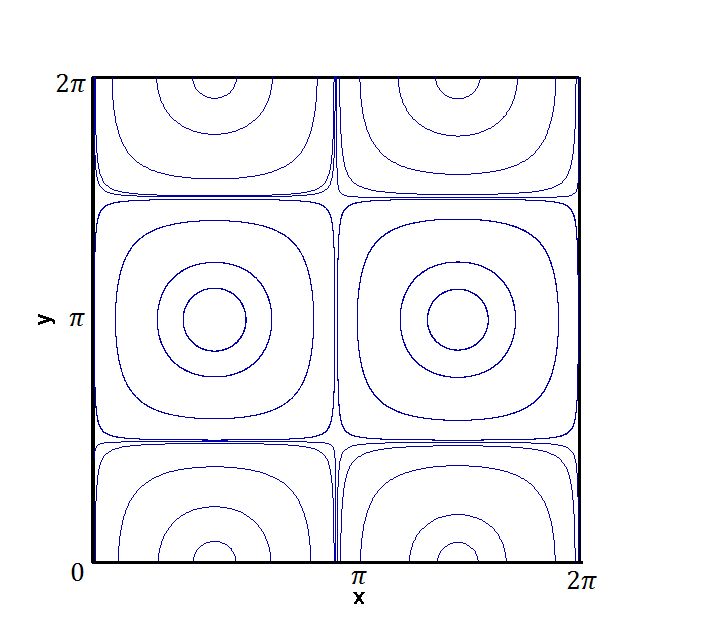}
        \label{fig:img2} }
  \caption{Streamlines (a) computed using the 2D symmetrized operator split cascaded LB forcing scheme and (b) obtained using the analytical solution for the four-rolls mill flow problem at $u_0=0.01$, $\nu=0.0011$ and $N=96$.}
\label{fig:fourrollmills2}
\end{figure}

Let us now consider a problem involving two-dimensional (2D), steady, fluid motion consisting of an array of counter-rotating vortices in a square domain of side $2\pi$ that is periodic in both $x$ and $y$ directions, driven by a spatially varying body force, i.e. $F_x=F_x(x,y)$ and $F_y=F_y(x,y)$. It is a modified form of the classical Taylor-Green vortex flow and has been used as a benchmark problem to test body force implementations in prior LBM studies (e.g.,~\cite{Silva2012,Derosis2017a}). The four-rolls fluid motion is established by imposing the following local body force components:
\begin{equation*}
F_x(x,y) = 2 \rho_0 \nu u_0 \sin x \sin y, \qquad F_y(x,y) = 2 \rho_0 \nu u_0 \cos x \cos y,
\end{equation*}
where $0\leq x,y \leq 2\pi$, $\rho_0$ is the reference density, $\nu$ is the kinematic viscosity, and $u_0$ is the velocity scale. A simplification of the
Navier-Stokes equations with the above local body force leads to the following analytical solution for the velocity field:
\begin{equation*}
u_x(x,y) = u_0 \sin x \sin y, \qquad u_y(x,y) = u_0 \cos x \cos y.
\end{equation*}
First, in order to validate the Strang splitting-based forcing scheme for the cascaded LBM, we consider $u_0=0.01$, $\rho_0=1.0$ and $\nu=0.0011$, and the square domain of side $2\pi$ is resolved by $N \times N$ mesh grids, where $N = 24, 48, 96, 192$. The mesh spacing $\Delta x$ then is given by $\Delta x = 2\pi/N$. Considering the convective scaling $\Delta x /\Delta t = c = 1$, the kinematic viscosity may be written as $\nu = \frac{1}{3}(\tau-\frac{1}{2})\Delta x$, where $\tau=1/\omega_4=1/\omega_5$. Figure~\ref{fig:fourrollmills1} shows the velocity field $u_y(x,y=\pi)$ computed using $N=96$ along the horizontal centerline of the domain and compared agains the analytical solution given above. Excellent agreement is seen.
\begin{figure}[htbp]
\centering
\advance\leftskip-1cm
\advance\rightskip 1cm
   {
        \includegraphics[scale=0.6] {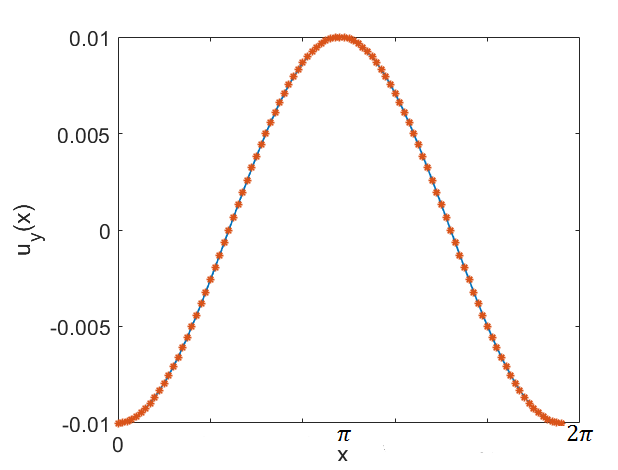}
        \label{fig:fourrollmills1} }
  \caption{Comparison of the computed and analytical vertical velocity profiles $u_y(x)$ at $y=\pi$ for the four-rolls mill flow problem at $u_0=0.01$, $\nu=0.0011$ and $N=96$. Here, line represents the analytical solution and the symbol refers to the numerical results obtained using the 2D symmetrized operator split cascaded LB forcing scheme.}
  \label{fig:fourrollmills1}
\end{figure}
Furthermore, Fig.~\ref{fig:fourrollmills2} presents the 2D computed and analytical results for the streamlines, which are in very good agreement with each other. Evidently, counter-rotating pairs of vortices are well reproduced by the present forcing scheme for the cascaded LBM based on Strang splitting.

\subsubsection*{Grid Convergence Study}
\begin{figure}[htbp]
\centering
\advance\leftskip-1cm
\advance\rightskip 1cm
   {
        \includegraphics[scale=.5] {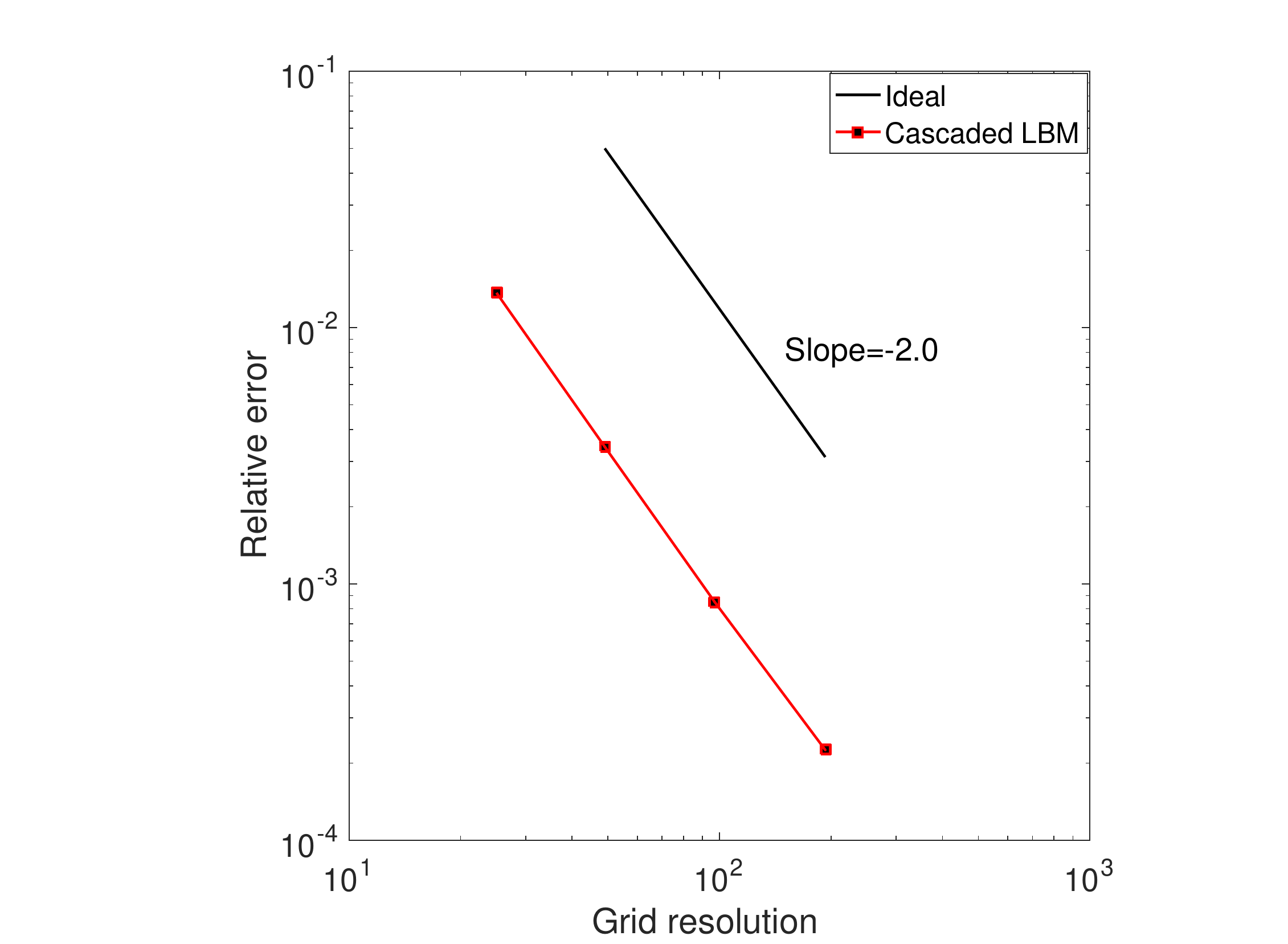}
   }
  \caption{Grid convergence for the four-rolls mill flow problem at $u_0=0.01$, $\nu=0.0011$ computed using the 2D symmetrized operator split cascaded LB forcing scheme under the convective scaling.}
\label{fig:fourrollmills3}
\end{figure}

In order to verify the higher order accuracy provided by the Strang splitting, i.e., $O(\Delta x^2)\sim O(\Delta t^2)$, we use the \emph{convective} or
\emph{acoustic} scaling to study the convergence rate of the present operator-split forcing formulation for different grid resolutions, rather than the \emph{diffusive} scaling considered earlier. Thus, we again use $u_0=0.01$, $\nu=0.0011$ and $N=24, 48, 96$ and $192$. By maintaining $\Delta x/\Delta t = c = 1$, for any pair of grid resolutions, $N_i \times N_i$ and $N_j \times N_j$, the corresponding relaxation parameters $\tau_i$ and
$\tau_j$, respectively, under the convective scaling are related by $\tau_j=\frac{1}{2}+(\tau_i-\frac{1}{2})\frac{N_j}{N_i}$. Figure~\ref{fig:fourrollmills3} illustrates rate of convergence using the relative error between the computed and analytical solution for the
$x$-component of the velocity field summed for the entire domain under the discrete $\ell_2$ norm (see Eq.~(\ref{eq:sca})) for the above four different grid resolutions. It can be seen that the relative error varies with the grid resolution in the log-log scale with a slope of $-2.0$. Hence, the present forcing scheme based on the Strang splitting for the cascaded LBM is second order accurate under the convective scaling. In other words, this test demonstrates second order accuracy in time, while the earlier test for Poiseuille flow under diffusive scaling in Fig.~\ref{fig:sca} is not.

\subsection{Thermal Couette Flow with Viscous Heat Dissipation}

\begin{figure}[htbp]
\centering
\advance\leftskip-1cm
\advance\rightskip-1cm
      {
        \includegraphics[scale=0.5] {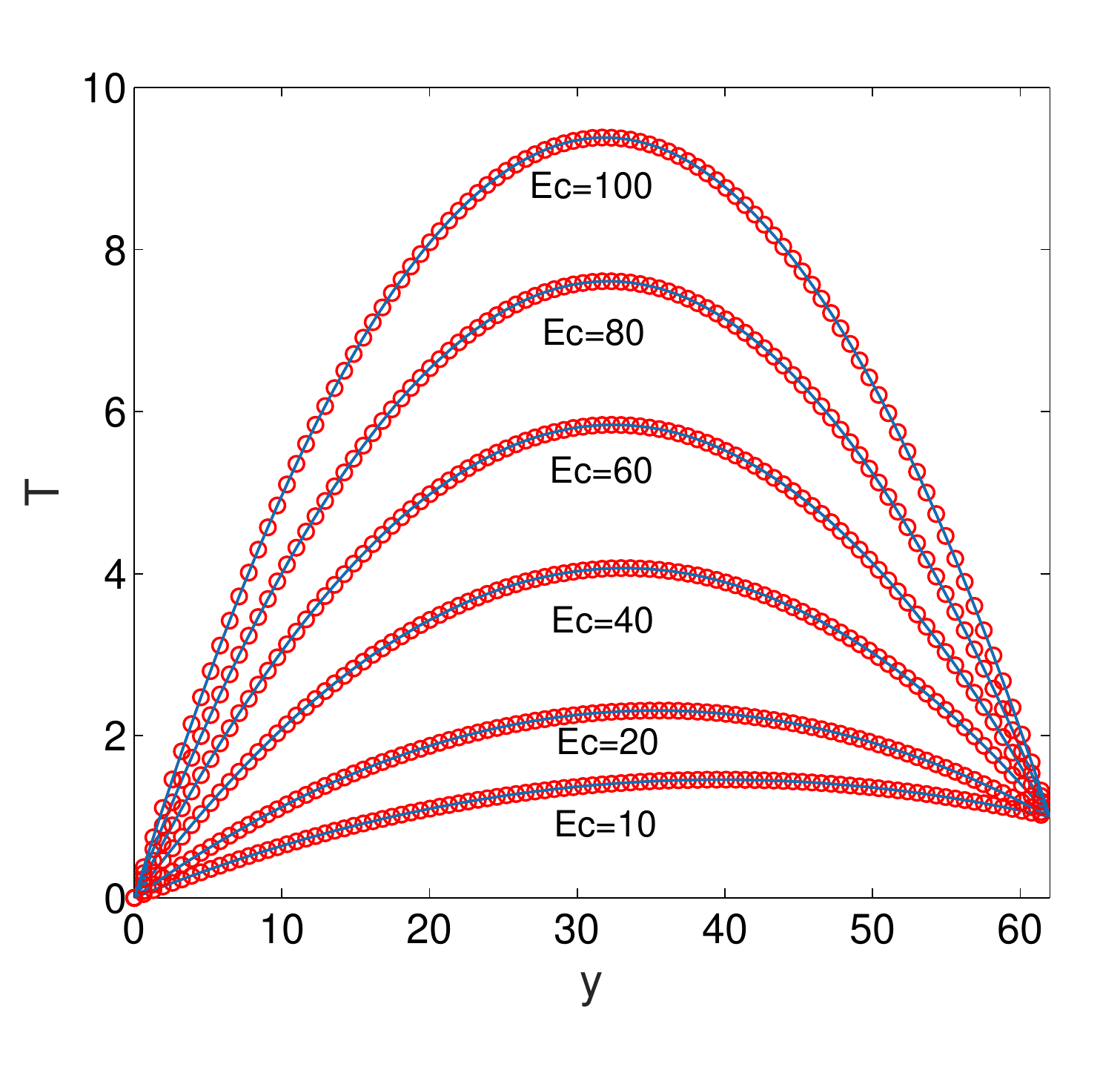}
        \label{fig:img2} }
        \caption{Comparison between numerical results of the temperature profile computed using the 2D symmetrized operator split cascaded LB source scheme for a passive scalar transport and the analytical solution for the thermal Couette flow for various values of the Eckert number $\mbox{Ec}$. Here, lines represent the analytical solution and symbols refer to the results obtained using the present numerical scheme.}
    \label{fig:cou}
\end{figure}

For the purpose of validating the symmetrized operator split cascaded source scheme for the solution of a scalar passive field represented by the CDE with a source term in Sec. 5, we perform the simulation of a thermal Couette flow with viscous heat dissipation. Here, the passive scalar field $\phi$ is the temperature $T$, which is evolved under a thermal diffusivity D, and modified by a source term $S_r$ due to viscous dissipation arising from the shear flow. For such a one-dimensional Couette flow, the top wall moves with a constant velocity $U_0$ in a horizontal direction, which is maintained at a higher temperature $T_H$ and the bottom wall is at a lower temperature $T_L$ and remains stationary. The scalar source term $S_r$ resulting from the viscous heat dissipation is given by
\begin{eqnarray}
S_r=\frac{2\nu}{C_v}(\tensr S: \tensr S),
\label{eq:pre1}
\end{eqnarray}
where $\tensr S=[\nabla u+{\nabla u}^T]/2$ is the strain rate tensor and $C_v$ is specific heat at constant volume. The source term due to the viscous heating $S_r$ in Eq.~(\ref{eq:pre1}) is obtained in simulations from the cascaded LB solution for the flow field presented in Sec.~\ref{app:Sec3}. In particular, the strain rate tensor $\tensr S$ in the cascaded LB formulation can be readily related to the second-order non-equilibrium moment components~(see e.g.,~\cite{Premnath2009for,ning2016numerical}). For example, $S_{xy}=\frac{1}{2}(\partial_xu_y+\partial_yu_x)=-\frac{3\omega_5}{2\rho_0}(\sum\limits_\alpha f_\alpha e_{\alpha x}e_{\alpha y}-\rho u_x u_y)$.
This problem has the following analytical solution for the temperature profile~\cite{Anderson1991}
\begin{eqnarray}
\frac{T-T_L}{T_H-T_L}=\frac{y}{H}+\frac{\mbox Pr \mbox Ec}{2}\frac{y}{H}\left(1-\frac{y}{H}\right),
\label{eq:cou}
\end{eqnarray}
where $\mbox Pr=\nu/D$ is the Prandtl number and $\mbox{Ec}=U_0^2/[C_v(T_H-T_L)]$ is the Eckert number. In Fig.~\ref{fig:cou}, the $\mbox{Pr}$ is fixed at $0.71$ while the $\mbox{Ec}$ varies from 10 to 100 and the domain is discretized with $3 \times 64$ lattice nodes. The velocity of the top wall $U_0$ is taken as $0.05$, the boundary temperature $T_L$ and $T_H$ are specified as $0.0$ and $1.0$, respectively, and the relaxation parameters $\tau$ and $\tau^\phi$ are chosen as $0.70$ and $0.782$, respectively. Computed results obtained using the symmetrized operator split cascaded source scheme are compared with the analytical solution given in Eq.~(\ref{eq:cou}). It is found that the numerical results are in excellent agreement with the analytical solution for various values of $\mbox{Ec}$, representing the source strength for this problem. In addition, the relative error between the computed results obtained using the Strang splitting-based source scheme and the analytical solution measured under the discrete $\ell_2$-norm (Eq.~(\ref{eq:sca})) for the simulation of the thermal Couette flow are reported in Table~\ref{table:relativeerrorthermalcouetteflow}.
\begin{table}[]
\centering
\caption{Relative error between the numerical results obtained using the 2D symmetrized operator split cascaded LB source scheme for a passive scalar transport and the analytical solution for the simulation of the thermal Couette flow at various Eckert numbers $\mbox{Ec}$.}
\label{table:relativeerrorthermalcouetteflow}
\begin{tabular}{|c|c|lll}
\cline{1-2}
Eckert number $\mbox{Ec}$   & Relative error    &  &  &  \\ \cline{1-2}
10                          & $2.840\times 10^{-5}$      &  &  &  \\ \cline{1-2}
20                          & $3.695\times 10^{-5}$      &  &  &  \\ \cline{1-2}
40                          & $4.317\times 10^{-5}$       &  &  &  \\ \cline{1-2}
60                          & $4.561\times 10^{-5}$       &  &  &  \\ \cline{1-2}
80                          & $4.691\times 10^{-5}$       &  &  &  \\ \cline{1-2}
100                         & $4.778\times 10^{-5}$        &  &  &  \\ \cline{1-2}
\end{tabular}
\end{table}

\section{\label{app:6}Summary and Conclusions}

Symmetrized operator split forcing schemes for flow simulations in 2D and 3D and a method for incorporating sources in a convection-diffusion transport of a scalar field using the cascaded lattice Boltzmann formulations are developed. They involve force/source implementation steps before and following the collision step each taking a half time step, and are consistent with the Strang splitting, which has second order rate of convergence by construction. The post-collision half source/forcing step is effectively implemented in terms of the change of moments at the zeroth/first order that is a function of the source/body force and the time step, and a normalization factor arising from the choice of the basis for moments for the lattice set considered. The implementation of the pre-collision half source/forcing step properly projects  the effects of the force/source to the higher order moments that undergo relaxation by collision and naturally eliminates the discrete effects. In contrast to the prior forcing schemes for the cascaded LB method that required using extra terms at different orders in the moment space and cumbersome lattice-dependent transformations to map them to the velocity space, the present symmetrized operator split forcing/source schemes result in a simpler formulation, with all the force/source related computations performed only in the moment space, which facilitates implementation. However, it may be noted that for efficient implementations of the LB algorithms, their performance on current hardware is limited entirely by memory bandwidth rather than by floating point operations, and the complexity of the aggregate collision operator (including forcing) does not affect performance. Comparisons of the numerical solutions obtained using the Strang splitting based forcing/source implementation methods for cascaded LB schemes against various benchmark solutions validate them for flow computations in both 2D and 3D as well as for the passive scalar transport with a local source. Furthermore, the numerical results demonstrate the second order accuracy for the convergence rate in time under the acoustic scaling of the symmetrized operator split forcing scheme.

\section*{Acknowledgements}
The authors would like to acknowledge the support of the US National Science Foundation (NSF) under Grant CBET-1705630.

\appendix

\section{\label{app:ChapmanEnskoganalysis}Strang Splitting Implementation of Body Forces in 3D Central Moment LB Method}
For the propose of illustration, we will consider the 3D central moment LB method using the three-dimensional, fifteen velocity (D3Q15)~\cite{Premnath2011three} lattice, but can be readily extended for other lattices such as the D3Q27 lattice. The components
of the particle velocity vectors along with the $\ket{1}$ vector (which is used to represent the zeroth moment with the distribution function)
for this lattice are
\begin{eqnarray}
&\ket{e_{x}} =\left(     0,     1,    -1,     0,     0,  0,0,1,-1, 1, -1,1,-1,1,-1 \right)^\dag, \nonumber\\
&\ket{e_{y}} =\left(     0,     0,    0,     1,     -1,  0,0,1,1, -1, -1,1,1,-1,-1 \right)^\dag,
\nonumber\\
&\ket{e_{z}} =\left(     0,     0,    0,     0,     0,  1,-1,1,1, 1, 1,-1,-1,-1,-1 \right)^\dag,
\nonumber\\
&\ket{1} =\left(     1,     1,    1,     1,     1,  1,1,1,1, 1, 1,1,1,1,1 \right)^\dag.
\label{eq:a1}
\end{eqnarray}
The corresponding linearly independent orthogonal basis vectors are given by ~\cite{Premnath2011three}
\begin{eqnarray}
&\ket{K_{0}}=\ket{1},\,
\ket{K_{1}}=\ket{e_{x}},\,
\ket{K_{2}}=\ket{e_{y}},\,
\ket{K_{3}}=\ket{e_{z}},\nonumber\\
&\ket{K_{4}}=\ket{e_{x}e_{y}},\,
\ket{K_{5}}=\ket{e_{x}e_{z}},\,
\ket{K_{6}}=\ket{e_{y}e_{z}},\nonumber\\
&\ket{K_{7}}=\ket{e_{x}^2-e_{y}^2},\,
\ket{K_{8}}=\ket{e_{x}^2+e_{y}^2+e_{z}^2}-3\ket{e_{z}^2},\,
\ket{K_{9}}=\ket{e_{x}^2+e_{y}^2+e_{z}^2}-2\ket{1},\nonumber\\
&\ket{K_{10}}=5\ket{e_{x}(e_{x}^2+e_{y}^2+e_{z}^2)}-13\ket{e_{x}},\nonumber\\
&\ket{K_{11}}=5\ket{e_{y}(e_{x}^2+e_{y}^2+e_{z}^2)}-13\ket{e_{y}},\,
\ket{K_{12}}=5\ket{e_{z}(e_{x}^2+e_{y}^2+e_{z}^2)}-13\ket{e_{z}},\nonumber\\
&\ket{K_{13}}=\ket{e_{x}e_{y}e_{z}},\nonumber\\
&\ket{K_{14}}=30\ket{e_{x}^2e_{y}^2+e_{x}^2e_{z}^2+e_{y}^2e_{z}^2}-40\ket{e_{x}^2+e_{y}^2+e_{z}^2}+32\ket{1}.
\label{eq:2A}
\end{eqnarray}
Then, the orthogonal matrix $\tensr K$ follows as
\begin{eqnarray}
\tensr{K}&=&\left[\ket{K_{0}},\ket{K_{1}},\ket{K_{2}},\ket{K_{3}},\ket{K_{4}},\ket{K_{5}},\ket{K_{6}},\ket{K_{7}},\ket{K_{8}}\right.\nonumber\\
           &&\left.\ket{K_{9}},\ket{K_{10}},\ket{K_{11}},\ket{K_{12}},\ket{K_{13}},\ket{K_{14}}\right],
\end{eqnarray}
which maps the change of moments under collisions back to the changes in the distribution functions.
The central moments and raw moments of the distribution function and its equilibrium of order ($m+n+p$) are defined, respectively, as
\begin{eqnarray}
\left( \begin{array}{l}
{{\hat \kappa }_{{x^m}{y^n}{z^p}}}\\
\hat \kappa _{{x^m}{y^n}{z^p}}^{eq}
\end{array} \right) = \sum\limits_\alpha  {\left( \begin{array}{l}
{f_\alpha }\\
f_\alpha ^{eq}\\
\end{array} \right)} {({e_{\alpha x}} - {u_x})^m}{({e_{\alpha y}} - {u_y})^n}{({e_{\alpha z}} - {u_z})^p},\label{eq:cen}
\end{eqnarray}
and
\begin{eqnarray}
\left( {\begin{array}{*{20}{l}}
{{{\hat \kappa }_{{x^m}{y^n}{z^p}}}}^{'}\\
{\hat {\kappa} _{{x^m}{y^n}{z^p}}^{eq'}}
\end{array}} \right) = \sum\limits_\alpha  {\left( {\begin{array}{*{20}{l}}
{{f_\alpha }}\\
{f_\alpha ^{eq}}
\end{array}} \right)} {e_{\alpha x}^m}{e_{\alpha y}^n}{e_{\alpha z}^p}.
\label{eq:g8}
\end{eqnarray}
The central moment equilibria used for the construction of the 3D cascaded collision operator for the D3Q15 lattice is presented in~\cite{Premnath2011three}. The collide and stream steps of the 3D cascaded method are formally represented in Eqs.~(\ref{eq:22a}) and~(\ref{eq:22b}), respectively. Owing to the mass and momentum being collision invariants, it follows that $\widehat{g_0}=\widehat{g_1}=\widehat{g_2}=\widehat{g_3}=0
$. For the non-conserved moments, the change of moments under cascaded collision are given by
\begin{eqnarray}
\widehat{g}_4&=&\frac{\omega_4}{8}\left[-\widehat{{\kappa}}_{xy}^{'}+\rho u_xu_y\right],\nonumber \\
\widehat{g}_5&=&\frac{\omega_5}{8}\left[-\widehat{{\kappa}}_{xz}^{'}+\rho u_xu_z\right],\nonumber \\
\widehat{g}_6&=&\frac{\omega_6}{8}\left[-\widehat{{\kappa}}_{yz}^{'}+\rho u_yu_z\right],\nonumber \\
\widehat{g}_7&=&\frac{\omega_7}{4}\left[-(\widehat{{\kappa}}_{xx}^{'}-\widehat{{\kappa}}_{yy}^{'})+\rho (u_x^2-u_y^2)\right],\nonumber \\
\widehat{g}_8&=&\frac{\omega_8}{12}\left[-(\widehat{{\kappa}}_{xx}^{'}+\widehat{{\kappa}}_{yy}^{'}-2\widehat{{\kappa}}_{zz}^{'})+\rho (u_x^2+u_y^2-2u_z^2)\right.\nonumber \\
 \widehat{g}_9&=&\frac{\omega_9}{18}\left[-(\widehat{{\kappa}}_{xx}^{'}+\widehat{{\kappa}}_{yy}^{'}+\widehat{{\kappa}}_{zz}^{'})+\rho (u_x^2+u_y^2+u_z^2)\right].\nonumber\\
\widehat{g}_{10}&=&\frac{\omega_{10}}{16}\left[-\widehat{{\kappa}}_{xyy}^{'}+2u_y\widehat{{\kappa}}_{xy}^{'}+u_x\widehat{{\kappa}}_{yy}^{'}-2\rho u_xu_y^2\right]+u_y\widehat{g}_4+\frac{1}{8}u_x(-\widehat{g}_7+\widehat{g}_8+3\widehat{g}_9),\nonumber \\
\widehat{g}_{11}&=&\frac{\omega_{11}}{16}\left[-\widehat{{\kappa}}_{xxy}^{'}+2u_x\widehat{{\kappa}}_{xy}^{'}+u_y\widehat{{\kappa}}_{xx}^{'}-2\rho u_x^2u_y\right]+u_x\widehat{g}_4+\frac{1}{8}u_y(\widehat{g}_7+\widehat{g}_8+3\widehat{g}_9),\nonumber \\
\widehat{g}_{12}&=&\frac{\omega_{12}}{16}\left[-\widehat{{\kappa}}_{xxz}^{'}+2u_x\widehat{{\kappa}}_{xz}^{'}+u_z\widehat{{\kappa}}_{xx}^{'}-2\rho u_x^2u_z\right]+u_x\widehat{g}_5+\frac{1}{8}u_z(\widehat{g}_7+\widehat{g}_8+3\widehat{g}_9),\nonumber \\
\widehat{g}_{13}&=&\frac{\omega_{13}}{8}\left[-\widehat{{\kappa}}_{xyz}^{'}+u_x\widehat{{\kappa}}_{yz}^{'}+u_y\widehat{{\kappa}}_{xz}^{'}+u_z\widehat{{\kappa}}_{xy}^{'}-2\rho u_xu_yu_z\right]
                    +u_z\widehat{g}_4+u_y\widehat{g}_5+u_x\widehat{g}_6,\nonumber  \\
\widehat{g}_{14}&=&\frac{\omega_{14}}{16}\left[-\widehat{{\kappa}}_{xxyy}^{'}+2u_x\widehat{{\kappa}}_{xyy}^{'}+2u_y\widehat{{\kappa}}_{xxy}^{'}
                    -u_x^2\widehat{{\kappa}}_{yy}^{'}-u_y^2\widehat{{\kappa}}_{xx}^{'}-4u_xu_y\widehat{{\kappa}}_{xy}^{'}\right.\nonumber\\
                    &&\left.+\widetilde{\widehat{\kappa}}_{xx}\widetilde{\widehat{\kappa}}_{yy}+3\rho u_x^2u_y^2\right]-2u_xu_y\widehat{g}_{4}+\frac{1}{8}(u_x^2-u_y^2)\widehat{g}_{7}\nonumber\\
                    &&+\frac{1}{8}(-u_x^2-u_y^2)\widehat{g}_{8}+\left(\frac{3}{8}(-u_x^2-u_y^2)-\frac{1}{2}\right)\widehat{g}_{9}+2u_x\widehat{g}_{10}+2u_y\widehat{g}_{11},\label{eq:6A}
\end{eqnarray}
The output velocity field $\bm u^0=(u_x^o,u_y^o,u_z^o)$ is obtained following the streaming step as
\begin{eqnarray}
 &\rho u^o_x=\sum_{\alpha=0}^{14} {f}_{\alpha} e_{\alpha x},\quad  \rho u^o_y=\sum_{\alpha=0}^{14} {f}_{\alpha} e_{\alpha y}\quad \rho u^o_z=\sum_{\alpha=0}^{14} {f}_{\alpha}  e_{\alpha z}.
\label{eq:7A}
\end{eqnarray}
As in the 2D case, the pre-collision forcing step $\tensr F^{1/2}$ involves the following update to the velocity field:
\begin{equation}
 u_x=\frac{1}{\rho}\left(\rho u^o_x+\frac{F_x}{2}\Delta t\right),\quad u_y=\frac{1}{\rho}\left(\rho u^o_y+\frac{F_y}{2}\Delta t\right),\quad u_z=\frac{1}{\rho}\left(\rho u^o_z+\frac{F_z}{2}\Delta t\right),
\label{eq:8A}
\end{equation}
which will be used in the determination of the cascaded collision based change of different moments, i.e. $\widehat{g}_{\beta}$, where $\beta=4,5,\cdots,14$ as given in Eq.~(\ref{eq:6A}). Analogously, the other post-collision step $\tensr {F}^{1/2}$ in the symmetrized operator splitting can be written as
\begin{equation}
 \rho u^p_x=\rho u_x+\frac{F_x}{2}\Delta t,\quad \rho u^p_y=\rho u_y+\frac{F_y}{2}\Delta t,\quad  \rho u^p_z=\rho u_z+\frac{F_z}{2}\Delta t,
\label{eq:9A}
\end{equation}
which, via Eq.~(\ref{eq:8A}), reads as
\begin{equation}
 \rho u^p_x=\rho u^o_x+{F_x}\Delta t,\quad \rho u^p_y=\rho u^o_y+{F_y}\Delta t, \quad  \rho u^p_z=\rho u^o_z+{F_z}\Delta t.
\label{eq:10A}
\end{equation}
In order to effectively introduce this effect into the 3D cascaded formulation, we take the first order moments of the post-collision distribution function $ f^{p}_{\alpha}=f_{\alpha}+(\tensr K\cdot \widehat {\mathbf g})_{\alpha}$, which yields
\begin{subequations}
\begin{eqnarray}
 &\rho u^p_x=\Sigma_{\alpha}f^p_{\alpha}e_{\alpha x}=\Sigma_{\alpha}f_{\alpha}e_{\alpha x}+\Sigma_{\beta}{\braket{{K_{\beta}}|{e_{x}}}} \widehat g_{\beta},
\label{eq:11aA}\\
 &\rho u^p_y=\Sigma_{\alpha}f^p_{\alpha}e_{\alpha y}=\Sigma_{\alpha}f_{\alpha}e_{\alpha y}+\Sigma_{\beta}{\braket{{K_{\beta}}|{e_{y}}}} \widehat g_{\beta},
\label{eq:11bA}\\
&\rho u^p_z=\Sigma_{\alpha}f^p_{\alpha}e_{\alpha z}=\Sigma_{\alpha}f_{\alpha}e_{\alpha z}+\Sigma_{\beta}{\braket{{K_{\beta}}|{e_{z}}}} \widehat g_{\beta}.
\label{eq:11cA}
\end{eqnarray}
\end{subequations}
Based on the orthogonal basis vectors $K_{\beta}$ given in Eq.~(\ref{eq:2A}), it follows that
\begin{equation}
 \Sigma_{\beta}{\braket{{K_{\beta}}|{e_{x}}}}g_{\beta}=10 \widehat {g}_1,\quad \Sigma_{\beta}{\braket{{K_{\beta}}|{e_{y}}}}g_{\beta}=10 \widehat {g}_2,\quad
 \Sigma_{\beta}{\braket{{K_{\beta}}|{e_{z}}}}g_{\beta}=10 \widehat {g}_3.
\label{eq:12A}
\end{equation}
Using Eqs.~(\ref{eq:11aA})-~(\ref{eq:11cA}) along with Eqs.~(\ref{eq:7A}) and ~(\ref{eq:12A}) and comparing with ~(\ref{eq:10A}), we obtain the following result for the change of first order moments due to the force field:
\begin{equation}
\widehat {g}_1=\frac{F_x}{10}\Delta t, \quad \widehat {g}_2=\frac{F_y}{10}\Delta t,\quad\widehat {g}_3=\frac{F_z}{10}\Delta t.
\label{eq:13A}
\end{equation}
Finally, using Eq.~(\ref{eq:13A}) and Eq.~(\ref{eq:6A}) for the change of moments under cascaded collision in $(\tensr K\cdot \widehat {\mathbf g})_{\alpha}$ and expanding it, we get the expressions for the post collision-distribution function, which read as
\begin{eqnarray}
{{f}}^p_{0}&=&{f}_{0}+\left[\widehat{g}_0-2\widehat{g}_9+32\widehat{g}_{14}\right], \nonumber\\
{{f}}^p
_{1}&=&{f}_{1}+\left[\widehat{g}_0+\widehat{g}_1+\widehat{g}_{7}+\widehat{g}_{8}-\widehat{g}_{9}-8\widehat{g}_{10}-8\widehat{g}_{14}\right],\nonumber\\
{{f}}^p
_{2}&=&{f}_{2}+\left[\widehat{g}_0-\widehat{g}_1+\widehat{g}_{7}+\widehat{g}_{8}-\widehat{g}_{9}+8\widehat{g}_{10}-8\widehat{g}_{14}\right],\nonumber\\
{{f}}
_{3}&=&{f}_{3}+\left[\widehat{g}_0+\widehat{g}_2-\widehat{g}_{7}+\widehat{g}_{8}-\widehat{g}_{9}-8\widehat{g}_{11}-8\widehat{g}_{14}\right],\nonumber\\
{{f}}^p
_{4}&=&{f}_{4}+\left[\widehat{g}_0-\widehat{g}_2-\widehat{g}_{7}+\widehat{g}_{8}-\widehat{g}_{9}+8\widehat{g}_{11}-8\widehat{g}_{14}\right],\nonumber\\
{{f}}^p
_{5}&=&{f}_{5}+\left[\widehat{g}_0+\widehat{g}_3-2\widehat{g}_{8}-\widehat{g}_{9}-8\widehat{g}_{12}-8\widehat{g}_{14}\right],\nonumber\\
{{f}}^p
_{6}&=&{f}_{6}+\left[\widehat{g}_0-\widehat{g}_3-2\widehat{g}_{8}-\widehat{g}_{9}+8\widehat{g}_{12}-8\widehat{g}_{14}\right],\nonumber\\
{{f}}^p
_{7}&=&{f}_{7}+\left[\widehat{g}_0+\widehat{g}_1+\widehat{g}_2+\widehat{g}_3+\widehat{g}_4+\widehat{g}_5+\widehat{g}_6+\widehat{g}_9+2\widehat{g}_{10}+2\widehat{g}_{11}+2\widehat{g}_{12}\right.\nonumber\\
                                &&\left.+\widehat{g}_{13}+2\widehat{g}_{14}\right],\nonumber\\
{{f}}^p
_{8}&=&{f}_{8}+\left[\widehat{g}_0-\widehat{g}_1+\widehat{g}_2+\widehat{g}_3-\widehat{g}_4-\widehat{g}_5+\widehat{g}_6+\widehat{g}_9-2\widehat{g}_{10}+2\widehat{g}_{11}+2\widehat{g}_{12}\right.\nonumber\\
                                &&\left.-\widehat{g}_{13}+2\widehat{g}_{14}\right],\nonumber\\
{{f}}^p
_{9}&=&{f}_{9}+\left[\widehat{g}_0+\widehat{g}_1-\widehat{g}_2+\widehat{g}_3-\widehat{g}_4+\widehat{g}_5-\widehat{g}_6+\widehat{g}_9+2\widehat{g}_{10}-2\widehat{g}_{11}+2\widehat{g}_{12}\right.\nonumber\\
                                &&\left.-\widehat{g}_{13}+2\widehat{g}_{14}\right],\nonumber\\
{{f}}^p
_{10}&=&{f}_{10}+\left[\widehat{g}_0-\widehat{g}_1-\widehat{g}_2+\widehat{g}_3+\widehat{g}_4-\widehat{g}_5-\widehat{g}_6+\widehat{g}_9-2\widehat{g}_{10}-2\widehat{g}_{11}+2\widehat{g}_{12}\right.\nonumber\\
                                &&\left.+\widehat{g}_{13}+2\widehat{g}_{14}\right],\nonumber\\
{{f}}^p
_{11}&=&{f}_{11}+\left[\widehat{g}_0+\widehat{g}_1+\widehat{g}_2-\widehat{g}_3+\widehat{g}_4-\widehat{g}_5-\widehat{g}_6+\widehat{g}_9+2\widehat{g}_{10}+2\widehat{g}_{11}-2\widehat{g}_{12}\right.\nonumber\\
                                &&\left.-\widehat{g}_{13}+2\widehat{g}_{14}\right],\nonumber\\
{{f}}^p
_{12}&=&{f}_{12}+\left[\widehat{g}_0-\widehat{g}_1+\widehat{g}_2-\widehat{g}_3-\widehat{g}_4+\widehat{g}_5-\widehat{g}_6+\widehat{g}_9-2\widehat{g}_{10}+2\widehat{g}_{11}-2\widehat{g}_{12}\right.\nonumber\\
                                &&\left.-\widehat{g}_{13}+2\widehat{g}_{14}\right],\nonumber\\
{{f}}^p
_{13}&=&{f}_{13}+\left[\widehat{g}_0+\widehat{g}_1-\widehat{g}_2-\widehat{g}_3-\widehat{g}_4-\widehat{g}_5+\widehat{g}_6+\widehat{g}_9+2\widehat{g}_{10}-2\widehat{g}_{11}-2\widehat{g}_{12}\right.\nonumber\\
                                &&\left.+\widehat{g}_{13}+2\widehat{g}_{14}\right],\nonumber\\
{{f}}^p
_{14}&=&{f}_{14}+\left[\widehat{g}_0-\widehat{g}_1-\widehat{g}_2-\widehat{g}_3+\widehat{g}_4+\widehat{g}_5+\widehat{g}_6+\widehat{g}_9-2\widehat{g}_{10}-2\widehat{g}_{11}-2\widehat{g}_{12}\right.\nonumber\\
                                &&\left.-\widehat{g}_{13}+2\widehat{g}_{14}\right].
\end{eqnarray}
The overall algorithmic sequence of steps for the 3D cascaded LB method with the operator split forcing implementation is similar to that presented in Sec.~4. Notice the significant simplification offered by the present 3D symmetrized operator split forcing scheme, when compared to that presented in~\cite{Premnath2011three}.

\end{document}